\documentclass[12pt]{article}
\usepackage[utf8]{inputenc}
\usepackage{amsmath}
\usepackage{mathtools}
\usepackage{booktabs}
\usepackage{cite}
\usepackage{amssymb}
\usepackage[titletoc,title]{appendix}
\usepackage{mathrsfs}
\usepackage{float}
\usepackage{multirow}
\usepackage{graphicx,caption,subcaption}
\usepackage[space]{grffile}
\usepackage{fullpage}
\usepackage{color}
\usepackage{soul}
\usepackage[dvipsnames]{xcolor}

\definecolor{darkblue}{rgb}{0.1,0.1,.7}
\usepackage[breaklinks=true,linktocpage=true,
colorlinks=true,urlcolor=darkblue,linkcolor=darkblue,
citecolor=darkblue,pdfpagelabels=true,hypertexnames=true,
plainpages=false,naturalnames=false,]{hyperref}

\usepackage{datetime}
\newdateformat{monthyeardate}{%
  \monthname[\THEMONTH], \THEYEAR}
\date{\monthyeardate\today}

\newcommand{\overbar}[1]{\mkern1.5mu\overline{\mkern-1.5mu#1\mkern-1.5mu}\mkern 1.5mu}

\begin{document}

\renewcommand{\arraystretch}{1.3}
\thispagestyle{empty}

{\hbox to\hsize{\vbox{\noindent December 2021 (revised)}}}

\noindent
\vskip2.0cm
\begin{center}

{\Large\bf Quintic constraints for ${\cal N}=2$ multiplets\\
\vglue.1in and complete SUSY breaking}

\vglue.3in

\hspace{-2pt}Yermek Aldabergenov,${}^{a,b,}$\footnote{yermek.a@chula.ac.th} Ignatios Antoniadis,${}^{c,d,}$\footnote{antoniad@lpthe.jussieu.fr} Auttakit Chatrabhuti,${}^{a,}$\footnote{auttakit.c@chula.ac.th} Hiroshi Isono${}^{a,}$\footnote{hiroshi.isono81@gmail.com}
\vglue.1in

${}^a$~{\it Department of Physics, Faculty of Science, Chulalongkorn University,\\ Phayathai Road, Pathumwan, Bangkok 10330, Thailand}\\
${}^b$~{\it Department of Theoretical and Nuclear Physics, 
Al-Farabi Kazakh National University,\\ 71 Al-Farabi Ave., Almaty 050040, Kazakhstan}\\
${}^c$~{\it Laboratoire de Physique Th\'eorique et Hautes Energies (LPTHE), Sorbonne Universit\'e,\\ CNRS, 4 Place Jussieu, 75005 Paris, France}\\
${}^d$~{\it Department of Mathematical Sciences, University of Liverpool,\\ Liverpool L69 7ZL, United Kingdom}
\vglue.1in

\end{center}

\vglue.3in

\begin{center}
{\Large\bf Abstract}
\vglue.2in
\end{center}

We study real fifth-order superfield constraint for ${\cal N}=2$ vector (and tensor) multiplet and derive most general solution describing complete supersymmetry breaking, and preserving a real scalar, two goldstini, and an abelian gauge field as low-energy degrees of freedom on which both supersymmetries are realized non-linearly. The surviving scalar is identified as an axion of a broken global abelian symmetry, while its scalar partner (saxion) is eliminated in terms of the goldstini. We provide an example of a UV model giving rise to the quintic constraint, and discuss the connection of this constraint and its solution to other known superfield constraints in ${\cal N}=2$ and ${\cal N}=1$ cases.

\newpage

\tableofcontents

\setcounter{footnote}{0}

\section{Introduction}

Introduced by Volkov and Akulov \cite{Volkov:1973ix}, non-linear realization of global supersymmetry (SUSY) is a useful tool for capturing low-energy behaviour of theories with spontaneously broken supersymmetry. Theories with non-linear ${\cal N}=1$ SUSY can be constructed off-shell by using constrained (nilpotent) chiral superfields \cite{Rocek:1978nb,Ivanov:1978mx,Lindstrom:1979kq,Ivanov:1982bpa,Samuel:1982uh,Casalbuoni:1988xh} 
(see also \cite{Komargodski:2009rz} for further discussion of nilpotent and orthogonal nilpotent superfields),
$\Phi^2=0$, and the resulting action can be related to the Volkov--Akulov theory by field redefinitions \cite{Rocek:1978nb,Lindstrom:1979kq,Kuzenko:2010ef}. 

Similar methods of constrained superfields can also be applied in ${\cal N}=2$ where there are two off-shell multiplets available: vector and tensor. In this case SUSY can be broken partially \cite{Hughes:1986dn,Hughes:1986fa,Ferrara:1995gu,Antoniadis:1995vb,Ferrara:1995xi,Bagger:1996wp,Ivanov:1997mt,Antoniadis:2017jsk} or completely. Bagger and Galperin \cite{Bagger:1996wp} showed that non-linear realization of partially broken ${\cal N}=2$ SUSY with abelian vector multiplet gives rise to supersymmetric Born--Infeld theory \cite{Born:1934gh,Cecotti:1986gb,Rocek:1997hi} with one linear and one non-linear and spontaneously broken SUSY. This construction has a counterpart in the ${\cal N}=2$ tensor multiplet case \cite{Bagger:1997pi,Gonzalez-Rey:1998vtf,Ambrosetti:2009za}. 

${\cal N}=2$ SUSY breaking in the non-linear limit can be conveniently described with the help of ${\cal N}=2$ supefields. Let us focus on ${\cal N}=2$ vector multiplet which can be embedded in an ${\cal N}=2$ superfield $X$ which is chiral w.r.t. both supersymmetries,
\begin{equation}
	\overbar D_{\dot\alpha}X=\overbar{\cal D}_{\dot\alpha}X=0~,\label{chiral-chiral_constr}
\end{equation}
where we call the two fermionic coordinates $\theta$ and $\vartheta$, and $D_\alpha$ and ${\cal D}_\alpha$ are the respective supercovariant derivatives (we use two-component spinor notation of \cite{Wess:1992cp}). The solution to \eqref{chiral-chiral_constr} that describes ${\cal N}=2$ vector multiplet reads
\begin{equation}
	X=\Phi+\sqrt{2}i\vartheta W+\vartheta^2\left(m-\tfrac{1}{4}\overbar{D}^2\overbar\Phi\right)~,
\end{equation}
where the ${\cal N}=1$ superfields $\Phi$ and $W_\alpha$ are chiral superfields in $\theta$-coordinate, and $W_\alpha$ is the field strength of a real ${\cal N}=1$ superfield $V$, defined as $W_\alpha\equiv -\tfrac{1}{4}\overbar{D}^2D_\alpha V$. Under the second SUSY these components transform as
\begin{equation}
	\delta_{\epsilon_2}\Phi=\sqrt{2}i\epsilon_2^\alpha W_\alpha~,~~~\delta_{\epsilon_2}W_\alpha=-\sqrt{2}i\epsilon_2^\alpha\left(m-\tfrac{1}{4}\overbar D^2\overbar\Phi\right)+\sqrt{2}\sigma^m_{\alpha\dot\alpha}\bar\epsilon_2^{\dot\alpha}\partial_m\Phi~,\label{intro_SUSY_transform}
\end{equation}
where $\epsilon_2$ is a constant transformation parameter. The real parameter $m$ is a magnetic FI term \cite{Antoniadis:1995vb} which can be used to partially break ${\cal N}=2$ supersymmetry. In the non-linear limit, this partial breaking can be described by applying quadratic nilpotency condition on $X$,
\begin{equation}
	X^2=0~.\label{intro_quad}
\end{equation}
The $\vartheta^2$-component of \eqref{intro_quad} yields the ${\cal N}=1$ superfield form of the constraint,
\begin{equation}
	\Phi\left(m-\tfrac{1}{4}\overbar D^2\overbar\Phi\right)=-\tfrac{1}{2}W^2~,\label{intro_quad_comp}
\end{equation}
where $W^2\equiv W^\alpha W_\alpha$. This leads to SUSY Born--Infeld theory once $\Phi$ is eliminated in the Lagrangian in terms of $W^2$. This can be done recursively thanks to the anticommuting nature of the fermionic superfield $W_\alpha$, which means $W^3=0$ (suppressing the indices). The first two $\vartheta$-components of \eqref{intro_quad} also show that $\Phi^2=\Phi W_\alpha=0$ which automatically holds if $\Phi$ is bilinear $\propto W^2$.

When both supersymmetries are non-linearly realized, this can be described by a cubic ${\cal N}=2$ superfield constraint introduced by Dudas, Ferrara, and Sagnotti (DFS) \cite{Dudas:2017sbi},~\footnote{See also \cite{Kuzenko:2017gsc} for further developments, and \cite{Cribiori:2016hdz} for the general discussion of $\cal N$-extended non-linear SUSY and its goldstino sector.}
\begin{equation}
	X^3=0~.\label{intro_cub}
\end{equation}
Here the $\vartheta^2$-component reads
\begin{equation}
\Phi^2\left(m-\tfrac{1}{4}\overbar D^2\overbar\Phi\right)=-\Phi W^2~.\label{intro_cub_comp}
\end{equation}
Of course, the partial breaking case with $X^2=0$ also solves $X^3=0$, or in the ${\cal N}=1$ language, the solution to \eqref{intro_quad_comp} solves Eq. \eqref{intro_cub_comp} as well. However there is a more general solution to \eqref{intro_cub_comp} that does not satisfy the constraint \eqref{intro_quad_comp} (and by extension Eq. \eqref{intro_quad}). This solution breaks both supersymmetries and eliminates the complex scalar in $\Phi$, rather than $\Phi$ itself, in terms of two goldstini associated with the two fermions (the chiral fermion in $\Phi$ and the gaugino).

In this work we show that there is an even higher-order and more general superfield constraint that can be imposed on ${\cal N}=2$ vector multiplet, if we consider real, rather than chiral, constraints on $X$ and $\overbar X$. This is somewhat similar to the ${\cal N}=1$ case, where in \cite{Aldabergenov:2021obf} it was shown that one can generalize the quadratic nilpotent superfield $\Phi^2=0$ to a cubic nilpotent superfield $(\Phi+\overbar\Phi)^3=0$, which is a weaker constraint and eliminates only the real scalar ${\rm Re}\,\Phi|_{\theta=0}$. This cubic superfield constraint is applicable to models where there is an axion (${\rm Im}\,\Phi|_{\theta=0}$) which is protected by an exact or approximate global shift symmetry (an analogous constraint exists when $\Phi$ transforms by a phase rotation under an abelian symmetry). In particular, in the case where the axion is the goldstone boson of spontaneous R-symmetry breaking, this constrained superfield could describe the low energy degrees of freedom of spontaneous supersymmetry and R-symmetry breaking (goldstino and R-axion). Its generalization to ${\cal N}=2$ is motivated by string theory D-branes due to the presence of a second bulk supersymmetry which is realized non-linearly on their world-volume.

The outline of our paper is the following. In Section \ref{sec_quintic_shift} we find that similar logic can be used to construct higher-order ${\cal N}=2$ superfield constraint, but the constraint in this case is of fifth order, $(X+\overbar X)^5=0$, due to the presence of two goldstini. We find the appropriate constraints and their solutions in two cases -- when the ${\cal N}=1$ superfield $\Phi$ (as a component of $X$) transforms under a $U(1)$ symmetry by a constant imaginary shift, and by a phase rotation. We show that in the limit of decoupling ${\cal N}=1$ vector multiplet, the quintic ${\cal N}=2$ constraint $(X+\overbar X)^5=0$ reduces to the cubic ${\cal N}=1$ constraint $(\Phi+\overbar\Phi)^3=0$ of \cite{Aldabergenov:2021obf}. In Section \ref{sec_tensor} we discuss the application of the quintic constraint to ${\cal N}=2$ tensor multiplet, and in Section \ref{sec_UV} we construct an example of a UV model that leads to the quintic constraint at low energies. We summarize our findings and conclude in Section \ref{sec_conclusion}. Full component form, to all orders in the fermions, of the quintic (shift-symmetric) constraint can be found in Appendix \hyperref[AppA]{A}, and more detailed study of the UV model can be found in Appendix \hyperref[AppB]{B}.

\section{Quintic constraint for ${\cal N}=2$ vector multiplet}\label{sec_quintic_shift}

As mentioned in Introduction, our goal is to study higher-order ${\cal N}=2$ superfield constraints which in general break both supersymmetries. In the case of a vector multiplet the constraint is expected to eliminate one of the two real scalars in terms of the goldstini, while preserving the other one (which we call the axion), protected by a global abelian symmetry, in analogy with the cubic constraints described in \cite{Aldabergenov:2021obf}.

Assuming that the abelian symmetry is realized as a shift symmetry, $X\rightarrow X+i\alpha$, where $\alpha$ is a real constant, we consider a real nilpotency constraint of the form
\begin{equation}
	(X+\overbar X)^n=0~,
\end{equation}
where $n$ is some positive integer.

If we expect both supersymmetries to be broken, the leading component of the constraint, which we call $\Phi|_{\theta=0}\equiv\phi$, must be a bilinear of the goldstini $\chi$ and $\lambda$,~\footnote{We denote the components of the ${\cal N}=1$ superfields as $\Phi=\{\phi,\chi_\alpha,F\}$ and $V=\{\lambda_\alpha,A_m,D\}$.}
\begin{equation}
\phi+\bar \phi\sim \chi^2+\lambda^2+\chi\lambda+\ldots+{\rm h.c.}\label{solution_form}
\end{equation}
Therefore we have
\begin{equation}
	(\phi+\bar \phi)^2\sim\chi\lambda\bar\chi\bar\lambda+\chi^2\lambda^2+\bar\chi^2\bar\lambda^2+\ldots~,~~~(\phi+\bar \phi)^4\sim\chi^2\lambda^2\bar\chi^2\bar\lambda^2~,
\end{equation}
and
\begin{equation}
	(\phi+\bar \phi)^5=0~.
\end{equation}
We then go back to the superfield level and impose this constraint on $X$,
\begin{equation}
	(X+\overbar X)^5=0~.\label{quintic_constr}
\end{equation}

\subsection{Component equation}

The $\vartheta^2\bar{\vartheta}^2$-component of the constraint \eqref{quintic_constr} reads
\begin{equation}
	\Phi_+^3\left(\tfrac{1}{8}\Phi_+\Box{\Phi_+}+A\right)-3\Phi_+^2B+3\Phi_+W^2\overbar W^2=0~,\label{quintic_subconstr}
\end{equation}
where $\Phi_\pm\equiv\Phi\pm\overbar\Phi$. $A$ and $B$ are the following functions of the ${\cal N}=1$ superfields,
\begin{align}
\begin{split}
	A &=2\left(m-\tfrac{1}{4}\overbar D^2\overbar\Phi\right)\left(m-\tfrac{1}{4}D^2\Phi\right)+\tfrac{1}{2}(\partial_m\Phi_-)^2-iW\sigma^m\partial_m\overbar W+i\partial_mW\sigma^m\overbar W~,\\
	B &=-\left(m-\tfrac{1}{4}D^2\Phi\right)W^2-\left(m-\tfrac{1}{4}\overbar D^2\overbar\Phi\right)\overbar W^2-iW\sigma^m\overbar W\partial_m\Phi_-~.
\end{split}
\end{align}

The component expansion of $\Phi$ and $W$ is
\begin{gather}
\begin{gathered}
	\Phi=\phi_1+i\phi_2+\sqrt{2}\theta\chi+\theta^2F~,\\
	W_\alpha=-i\lambda_\alpha+(\delta^{\beta}_\alpha D+\tfrac{i}{2}\sigma^m_{\alpha\dot\alpha}\overbar\sigma^{n\dot\alpha\beta}F_{mn})\theta^\beta+\theta^2\sigma^m_{\alpha\dot\alpha}\partial_m\bar\lambda^{\dot\alpha}~.\label{W_Phi_exp}
\end{gathered}
\end{gather}
Next we extract the $\theta^2\bar\theta^2$-component of Eq. \eqref{quintic_subconstr}, which is our master equation:
\begin{align}\label{master_eq}
\begin{split}
-\tfrac{2}{3}\phi_1^4\Box^2\phi_1+\phi_1^3\left(P+\tfrac{40}{3}\Box\phi_1\Box\phi_1+\tfrac{16}{3}\partial^{mn}\phi_1\partial_{mn}\phi_1\right)+\phi_1^2\left(J_1+J_2^{mn}\partial_{mn}\phi_1\right)&\\+\phi_1\left(I_1+I_2^{mn}\partial_{mn}\phi_1\right)+H_1+H_2^{mn}\partial_{mn}\phi_1 &=0~,
\end{split}
\end{align}
where $\partial_{mn}\equiv\partial_m\partial_n$. We introduced $P,H,I,J$ as (real) functions of the independent fields $\phi_2,\chi,\lambda,F_{mn},D,F$ and their spacetime derivatives. Here it is sufficient to write down these functions at the leading order in the fermions $\chi$ and $\lambda$, while their full expressions can be found in Appendix \hyperref[AppA]{A}.

$P$ starts with bosonic terms (meaning without fermions or their derivatives) and is defined as (this is a full expression for $P$)
\begin{align}	
\begin{split}
3P &\equiv 16\Box\phi_2\Box\phi_2+16\partial_{mn}\phi_2\partial^{mn}\phi_2+4\Box(\Omega+i\chi\sigma^m\partial_m\bar\chi+i\lambda\sigma^m\partial_m\bar\lambda)-32\partial_mF\partial^m\overbar F\\
&-16\partial_mD\partial^mD+8\partial^mF_{mn}\partial_kF^{kn}+4\partial_kF_{mn}\partial^kF^{mn}+4i\tilde{F}_{mn}\Box F^{mn}\\
&-16i\Box\chi\sigma^m\partial_m\bar\chi-16i\Box\lambda\sigma^m\partial_m\bar\lambda-16i\partial_{mn}\chi\sigma^m\partial^n\bar\chi-16i\partial_{mn}\lambda\sigma^m\partial^n\bar\lambda+{\rm h.c.}~
\end{split}
\end{align}
Here for convenience we introduce the shorthands
\begin{gather}\label{Omega_def}
\begin{gathered}
	\Omega\equiv D^2+2\bar fF-\tfrac{1}{2}F\cdot F-\tfrac{i}{2}F\cdot\tilde F-2\partial\phi_2\partial\phi_2-2i\chi\sigma^m\partial_m\bar\chi-2i\lambda\sigma^m\partial_m\bar\lambda~,\\
	f\equiv F+m~,~~~\tilde F_{mn}\equiv\tfrac{1}{2}\epsilon_{mnkl}F^{kl}~,~~~F^\pm_{mn}\equiv F_{mn}\pm i\tilde F_{mn}~,
\end{gathered}
\end{gather}
and use the notation $F\cdot F\equiv F_{mn}F^{mn}$ and $\partial A\partial B\equiv \partial_mA\partial^mB$.

Next, for $J_1$ and $J_2^{mn}$ we have
\begin{align}
\begin{split}
	J_1 &\equiv 8i\Omega\Box\phi_2+16\tilde F_{mn}\partial_kF^{nk}\partial^m\phi_2-2\Box(\chi^2\overbar F+\lambda^2f+\chi\sigma^m\bar\chi\partial_m\phi_2+\lambda\sigma^m\bar\lambda\partial_m\phi_2)\\
	&-8i\partial^m(2\bar fF-i\chi\sigma^n\partial_n\bar\chi-i\lambda\sigma^n\partial_n\bar\lambda)\partial_m\phi_2+16\sqrt{2}i\partial_m\chi\sigma^{mn}\partial_n\lambda D\\
&+8(2\partial^m\chi\sigma_n\partial_m\bar\chi+2\partial^m\lambda\sigma_n\partial_m\bar\lambda-i\epsilon_{mnkl}\partial^m\chi\sigma^k\partial^l\bar\chi-i\epsilon_{mnkl}\partial^m\lambda\sigma^k\partial^l\bar\lambda)\partial^n\phi_2\\
&-8\partial_m\chi\sigma^m\overbar\sigma^n\partial_n\chi\bar f-8\partial_m\lambda\sigma^m\overbar\sigma^n\partial_n\lambda F-4\sqrt{2}i\partial^m\chi\partial^n\lambda(\eta_{mn}D+2iF^+_{mn})\\
&-4\sqrt{2}(\partial^m\chi\sigma^{nk}\partial_k\lambda-\partial^m\lambda\sigma^{nk}\partial_k\chi)(F^+_{mn}+2i\tilde F_{mn})+{\rm h.c.}+\ldots~,
\end{split}\\[5pt]
\begin{split}
J_2^{mn} &\equiv 8\eta^{mn}(\tfrac{3}{2}\Omega+\tfrac{3}{2}\overbar\Omega+4\partial\phi_2\partial\phi_2+m^2+F\cdot F)+16\eta_{kl}F^{mk}F^{ln}-32\partial^m\phi_2\partial^n\phi_2+\ldots~
\end{split}
\end{align}
As can be seen both $J_1$ and $J_2^{mn}$ start with bosonic terms. The ellipsis denotes terms with more non-derivative fermions.

For $I_1$ and $I_2^{mn}$ we find
\begin{align}
\begin{split}
	I_1 &\equiv 4\Omega\overbar\Omega-16(m^2+F\cdot F)\partial\phi_2\partial\phi_2-32F^{mn}F_{nk}\partial^k\phi_2\partial_m\phi_2+\ldots~,
\end{split}\\[5pt]
\begin{split}
I_2^{mn} &\equiv -4\eta^{mn}\Big[\chi^2(\overbar F+2\bar f)+\lambda^2(f+2F)+(\chi\sigma^k\bar\chi+\lambda\sigma^k\bar\lambda)\partial_k\phi_2\\
&\hspace{9cm}-3\sqrt{2}i\chi\lambda D+\sqrt{2}\chi\sigma^{kl}\lambda F_{kl}\Big]\\
&-8(\chi\sigma^m\bar\chi+\lambda\sigma^m\bar\lambda)\partial^n\phi_2+16\sqrt{2}\chi\sigma^{ml}\lambda F_{lk}\eta^{kn}+{\rm h.c.}~,
\end{split}
\end{align}
where $I_1$ starts with bosonic terms, while $I_2^{mn}$ is bilinear in $\chi$ and $\lambda$.

Finally, $H_1$ and $H_2^{mn}$ are given by
\begin{align}
\begin{split}
	H_1 &\equiv -\chi^2(2\bar f\overbar\Omega+4m\partial\phi_2\partial\phi_2)-\lambda^2(2F\overbar\Omega-4 m\partial\phi_2\partial\phi_2)\\
	&-\chi\sigma^m\bar\chi\partial^n\phi_2\Big[\eta_{mn}(2D^2+F\cdot F+4|f|^2-4\partial\phi_2\partial\phi_2)+4D\tilde F_{mn}+4\eta^{kl}F_{ml}F_{kn}\Big]\\
	&-\lambda\sigma^m\bar\lambda\partial^n\phi_2\Big[\eta_{mn}(2D^2+F\cdot F+4|F|^2-4\partial\phi_2\partial\phi_2)-4D\tilde F_{mn}+4\eta^{kl}F_{ml}F_{kn})\Big]\\
	&+2\sqrt{2}i\overbar\Omega(\chi\lambda D+i\chi\sigma^{mn}\lambda F_{mn})-8\sqrt{2}\chi\sigma^{mn}\lambda(F_{mn}\partial\phi_2\partial\phi_2+2F_{nk}\partial^k\phi_2\partial_m\phi_2)\\
	&+4\sqrt{2}i\chi\sigma^m\bar\lambda\partial^n\phi_2\left[\overbar F(\eta_{mn}D-iF^+_{mn})-\bar f(\eta_{mn}D-iF^-_{mn})\right]+{\rm h.c.}+\ldots~,
\end{split}\\[5pt]
	H_2^{mn} &\equiv \eta^{mn}(\chi^2\bar\chi^2+\lambda^2\bar\lambda^2+2\chi^2\lambda^2+2\bar\chi^2\bar\lambda^2)-4\chi\sigma^m\bar\chi\lambda\sigma^n\bar\lambda~.
\end{align}
Here $H_1$ is at least bilinear and $H_2^{mn}$ is quadrilinear.

It is easy to check that the constraint \eqref{master_eq} is invariant under the discrete $R$-symmetry (see e.g. \cite{Fayet:1975yi,Dudas:2017sbi})
\begin{equation}
	\chi\rightarrow i\lambda~,~~~\lambda\rightarrow\pm i\chi~,~~~F\rightarrow -\bar f~,~~~D\rightarrow\mp D~,~~~F_{mn}\rightarrow\pm F_{mn}~.
\end{equation}

As can be seen from Eq. \eqref{master_eq}, the solution $\phi_1$ has the form $\sim (H_1+\ldots)/I_1$ which is well-defined only if the bosonic part of $I_1$ is non-vanishing,
\begin{equation}
	I_1|_{\rm bos}=4\Omega\overbar\Omega-16(m^2+F\cdot F)\partial\phi_2\partial\phi_2-32F^{mn}F_{nk}\partial^k\phi_2\partial_m\phi_2\neq 0~.
\end{equation}
In particular this means that at the vacuum we have $\langle\Omega\rangle =\langle D^2+2(\overbar F+m)F\rangle\neq 0$, i.e. at least one of the auxiliary fields $F$ and $D$ must be non-vanishing. Then, by looking at SUSY transformation of the fermions at the vacuum,
\begin{align}
	\langle\delta_{\epsilon}\chi\rangle &=\sqrt{2}\epsilon_1\langle F\rangle+i\epsilon_2\langle D\rangle~,\\
	\langle\delta_{\epsilon}\lambda\rangle &=i\epsilon_1\langle D\rangle+\sqrt{2}i\epsilon_2\langle\overbar F+m\rangle~,\label{lambda_SUSY_transform}
\end{align}
we conclude that both supersymmetries must be broken, while the parameter $m$ sets the hierarchy between the two breaking scales.

\subsection{Leading-order solution and special case $X^3=0$}

At the leading (bilinear) order Eq. \eqref{master_eq} can be readily solved as 
\begin{equation}
	\phi_1=-(H_1/I_1)|_2+\ldots~,\label{bilinear_sol}
\end{equation}
where we introduce the notation $|_n$ which means extracting terms that are at most $n^{\rm th}$ order in $\chi,\bar\chi,\lambda,\bar\lambda$. For example $|_2$ extracts terms that are at most bilinear such as $\chi^2$, $\lambda^2$, $\chi\lambda$, $\chi\sigma^m\bar\lambda$, etc.

Since there is a particular solution to the constraint $(X+\overbar X)^5=0$ which satisfies $X^3=0$, we should be able to reproduce the DFS solution \cite{Dudas:2017sbi} which eliminates both real scalars. The lowest component of $X^3=0$ implies that both $\phi_1$ and $\phi_2$ are bilinear, and therefore the terms in \eqref{bilinear_sol} containing $\partial_m\phi_2$ include three or more non-derivative fermions, and can be ignored at the leading order. This yields
\begin{equation}
	\phi_1=\frac{1}{2}(\phi+\bar\phi)=\frac{\chi^2\bar f+\lambda^2 F-\sqrt{2}i(\chi\lambda D+i\chi\sigma^{mn}\lambda F_{mn})}{2D^2+4\bar fF-F_{mn}F^{mn}-iF_{mn}\tilde F^{mn}}+{\rm h.c.}+\ldots~,
\end{equation}
which agrees with the result of \cite{Dudas:2017sbi}. Without the additional constraint $X^3=0$, the solution \eqref{bilinear_sol} includes $\phi_2$ as a physical real scalar.

\subsection{Full solution}

Let us now solve the constraint up to eighth-order in $\chi,\bar\chi,\lambda,\bar\lambda$, since all higher-order terms identically vanish. As we showed earlier, the fifth power of bilinear functions vanish, i.e.
\begin{equation}
\phi_1^5=(I^{mn}_2)^5=H_1^5=0~.
\end{equation}
Since $H_2^{mn}$ is quadrilinear we have $(H^{mn}_2)^3=0$. Of course any product of these functions containing more than eight fermions also vanishes, e.g. $H_1^3H_2^{mn}=0$. This will help to solve the constraint \eqref{master_eq} by iteration.

It is convenient to rewrite the constraint \eqref{master_eq} as
\begin{equation}
	-\tfrac{2}{3}\phi_1^4\Box^2\phi_1+\phi_1^3G_3+\phi_1^2G_2+\phi_1G_1+G_0=0~,\label{master_eq_G}
\end{equation}
where
\begin{align}
\begin{split}
G_0 &=H_1+H_2^{mn}\partial_{mn}\phi_1~,\\
G_1 &=I_1+I_2^{mn}\partial_{mn}\phi_1~,\\
G_2 &=J_1+J_2^{mn}\partial_{mn}\phi_1~,\\
G_3 &=P+\tfrac{40}{3}\Box\phi_1\Box\phi_1+\tfrac{16}{3}\partial^{mn}\phi_1\partial_{mn}\phi_1~.\label{eq_G_HIJP}
\end{split}
\end{align}
Because $(I_1,J_1,J_2^{mn},P)$ start with bosonic terms, $H_1$ and $I_2^{mn}$ with bilinear, and $H_2^{mn}$ with fourth-order terms, it can be seen that $G_0$ is at least bilinear, while $G_{1,2,3}$ start from bosonic terms.

From \eqref{master_eq_G} we have
\begin{equation}
	\phi_1=-\frac{1}{G_1}\left(G_0+\phi_1^2G_2+\phi_1^3G_3-\tfrac{2}{3}\phi_1^4\Box^2\phi_1\right)~.\label{master_eq_phi_G}
\end{equation}
As a first step we eliminate $\phi_1^2$, $\phi_1^3$, and $\phi_1^4$ on the RHS in terms of $G_{0,1,2,3}$. Squaring \eqref{master_eq_phi_G} we obtain,
\begin{equation}
	\phi_1^2=\frac{G_0^2+2\phi_1^3G_0G_3+\phi_1^4G_2^2}{G_1^2-2G_0G_2}~.\label{eq_phi^2}
\end{equation}
Further multiplying by \eqref{master_eq_phi_G} leads to
\begin{equation}
	\phi_1^3=-\frac{G_0^3(G_1^2-G_0G_2)}{G_1(G_1^2-2G_0G_2)^2}~,~~~\phi_1^4=\frac{G_0^4}{G_1^4}~.\label{eq_phi^3}
\end{equation}
Using Eqs. \eqref{eq_phi^2} and \eqref{eq_phi^3} in \eqref{master_eq_phi_G} we can write
\begin{equation}
	\phi_1=-\frac{G_0}{G_1}-\frac{G_0^2G_2}{G_1^3}+\frac{G_0^3G_3}{G_1^4}-\frac{2G_0^3G_2^2}{G_1^5}+\frac{5G_0^4G_2G_3}{G_1^6}-\frac{5G_0^4G_2^3}{G_0^7}+\frac{2G_0^4}{3G_1^5}\Box^2\phi_1~,\label{eq_phi_G_deriv}
\end{equation}
where $\phi_1$ enters the RHS only through its derivatives. 

The next step is to eliminate these derivatives, namely $\partial_{mn}\phi_1$ and $\Box^2\phi_1$, in terms of the independent fields contained in the functions $H,I,J,P$. To do so we notice that in \eqref{eq_phi_G_deriv} (or \eqref{master_eq_phi_G}), the second derivative $\partial_{mn}\phi_1$ always multiplies at least four goldstini, which means we must find $\partial_{mn}\phi_1$ up to fourth-order terms. As for $\Box^2\phi_1$, it suffices to obtain its bosonic part $\Box^2\phi_1|_0$ since it multiplies $\phi_1^4\sim G_0^4\sim \chi^2\bar\chi^2\lambda^2\bar\lambda^2$.

It is useful to find the leading-order term of $\partial_{mn}\phi_1$. By applying $\partial_{mn}$ to \eqref{eq_phi_G_deriv} we obtain a simple expression $\partial_{mn}\phi_1|_0=-(\partial_{mn}H_1)/I_1$.~\footnote{In our notation, $|_n$ in $\partial_{mn}\phi_1|_n$ is applied after the derivatives are taken.} Now we can derive $\Box^2\phi_1|_0$ by applying $\Box^2$ to \eqref{eq_phi_G_deriv} and extracting the bosonic terms. We get
\begin{equation}
	\Box^2\phi_1|_0=-\Box^2\frac{H_1+H_2^{mn}\partial_{mn}\phi_1}{I_1+I_2^{mn}\partial_{mn}\phi_1}-\frac{\Box^2H_1^2}{I_1^3}(J_1+J_2^{mn}\partial_{mn}\phi_1)~.\label{eq_Box^2_1}
\end{equation}
Since here we ignore terms proportional to goldstini, we can use $\partial_{mn}\phi_1|_0=-(\partial_{mn}H_1)/I_1$ on the RHS. Taking this into account and Taylor-expanding the first term of \eqref{eq_Box^2_1} in $I_2^{mn}$ (recall that it is at least bilinear in goldstini) we arrive at the final expression
\begin{equation}
	\Box^2\phi_1|_0=\frac{1}{I_1^2}\Box^2H_2^{mn}\partial_{mn}H_1-\Box^2\left(\frac{H_1}{I_1}+\frac{H_1}{I_1^3}I_2^{mn}\partial_{mn}H_1\right)-\frac{\Box^2H_1^2}{I_1^4}(I_1J_1-J_2^{mn}\partial_{mn}H_1)~.\label{eq_Box^2_final}
\end{equation}

Before considering $\partial_{mn}\phi_1|_4$ let us obtain an expression for $\partial_{mn}\phi_1$ up to bilinear terms. From Eq. \eqref{eq_phi_G_deriv} we have
\begin{align}\label{eq_Box_phi_bi}
\begin{split}
	\partial_{mn}\phi_1|_2 &=-\partial_{mn}\left(\frac{G_0}{G_1}+\frac{G_0^2G_2}{G_1^3}\right)\\
	&=-\partial_{mn}\left(\frac{H_1}{I_1}-\frac{1}{I_1^2}H_2^{kl}\partial_{kl}H_1+\frac{H_1}{I_1^3}I_2^{kl}\partial_{kl}H_1\right)-\frac{\partial_{mn}H_1^2}{I_1^4}(I_1J_1-J_2^{kl}\partial_{kl}H_1)~,
\end{split}
\end{align}
where we used $\partial_{mn}\phi_1|_0=-(\partial_{mn}H_1)/I_1$ and Taylor-expansion in $I_2^{mn}$.

Now we are in a position to derive $\partial_{mn}\phi_1$ up to fourth order in goldstini. Again, from \eqref{eq_phi_G_deriv} we have
\begin{equation}\label{eq_Box_phi_quad}
	\partial_{mn}\phi_1|_4=-\partial_{mn}\left(\frac{G_0}{G_1}+\frac{G_0^2G_2}{G_1^3}-\frac{G_0^3G_3}{G_1^4}+\frac{2G_0^3G_2^2}{G_1^5}\right)~.
\end{equation}
From the definitions \eqref{eq_G_HIJP} of the functions $G_{0,1,2,3}$ it can be seen that on the RHS of Eq. \eqref{eq_Box_phi_quad} the derivative $\partial_{mn}\phi_1$ is needed only up to bilinear terms, as higher-order terms do not contribute to $\partial_{mn}\phi_1|_4$. Expanding the $G$-functions, we have
\begin{align}\label{eq_Box_phi_quad_final}
\begin{split}
	\partial_{mn}\phi_1|_4= &-\partial_{mn}\bigg\{\frac{1}{I_1^5}(H_1+H_2^{kl}\partial_{kl}\phi_1|_2)\bigg[I_1^3(I_1-I_2^{pq}\partial_{pq}\phi_1|_2)+(I_2^{pq}\partial_{pq}H_1)^2\bigg]\\
	&+(J_1+J_2^{kl}\partial_{kl}\phi_1|_2)(H_1I_1^2-2I_1H_2^{pq}\partial_{pq}H_1+3H_1I_2^{pq}\partial_{pq}H_1)\bigg\}\\
	&+\frac{\partial_{mn}H_1^3}{I_1^6}\left(I_1^2P+\tfrac{40}{3}\Box H_1\Box H_1+\tfrac{16}{3}\partial^{kl}H_1\partial_{kl}H_1\right)-\frac{2\partial_{mn}H_1^3}{I_1^7}(I_1J_1-J_2^{kl}\partial_{kl}H_1)^2~.
\end{split}
\end{align}

Knowing Eqs. \eqref{eq_Box^2_final}, \eqref{eq_Box_phi_bi}, and \eqref{eq_Box_phi_quad_final} we can write down the full solution to the quintic constraint by inserting these into \eqref{eq_phi_G_deriv}. We obtain
\begin{align}
\begin{split}\label{full_sol}
	\phi_1= &-\frac{H_1+H_2^{mn}\partial_{mn}\phi_1|_4}{I_1+I_2^{kl}\partial_{kl}\phi_1|_4}-\frac{(H_1+H_2^{mn}\partial_{mn}\phi_1|_2)^2(J_1+J_2^{kl}\partial_{kl}\phi_1|_4)}{I_1+I_2^{pq}\partial_{pq}\phi_1|_2}\\
	&+\frac{H_1^2(H_1I_1-3H_2^{mn}\partial_{mn}H_1)\left(P+\tfrac{40}{3}(\Box\phi_1|_2)^2+\tfrac{16}{3}\partial^{kl}\phi_1|_2\partial_{kl}\phi_1|_2\right)}{I_1^3(I_1^2-4I_2^{pq}\partial_{pq}H_1)}\\
	&-\frac{2H_1^2(H_1I_1-3H_2^{mn}\partial_{mn}H_1)(J_1+J_2^{kl}\partial_{kl}\phi_1|_2)^2}{I_1^4(I_1^2-5I_2^{pq}\partial_{pq}H_1)}\\
	&+\frac{5H_1^4}{I_1^7}(I_1J_1-J_2^{mn}\partial_{mn}H_1)\left(P+\frac{40}{3I_1^2}\Box H_1\Box H_1+\frac{16}{3I_1^2}\partial^{kl}H_1\partial_{kl}H_1\right)\\
	&-\frac{5H_1^4}{I_1^{10}}(I_1J_1-J_2^{mn}\partial_{mn}H_1)^3+\frac{2H_1^4}{3I_1^5}\Box^2\phi_1|_0~.
\end{split}	
\end{align}
All the leading-order (bilinear) terms come from $-H_1/I_1$, while the rest of the expression is higher-order, containing up to eight (non-derivative) fermions.

\subsection{$U(1)$ as a phase symmetry}

The constraint $(X+\overbar X)^5=0$ is applicable in situations where $X$ transforms by imaginary shift under a global (exact or approximate) $U(1)$, and the corresponding axion is identified with the imaginary scalar component of $X$.

A $U(1)$ symmetry can also act on the ${\cal N}=1$ chiral superfield as a phase rotation. If it is not an $R$-symmetry (fermionic coordinates are inert), $m$ is forced to vanish since it will introduce non-invariant terms in the action, and the three components of $X$ transform as,
\begin{equation}
	\Phi\rightarrow\Phi e^{i\alpha}~,~~~W_\alpha\rightarrow W_\alpha~,~~~\overbar D^2\overbar\Phi\rightarrow\overbar D^2\overbar\Phi\,e^{-i\alpha}~.
\end{equation}
In contrast, if the $U(1)$ is an $R$-symmetry, it rotates the fermionic coordinates $\{\theta,\vartheta\}\rightarrow \{\theta,\vartheta\}e^{i\alpha/2}$, and therefore we have $W_\alpha\rightarrow W_\alpha\,e^{i\alpha/2}$ and $\overbar D^2\overbar \Phi\rightarrow \overbar D^2\overbar \Phi$. As the result, $X$ transforms homogeneously as $X\rightarrow Xe^{i\alpha}$, while the magnetic parameter $m$ is not required to vanish.

Regardless of whether the $U(1)$ phase symmetry is $R$-symmetry or not, we can impose the following invariant constraint at ${\cal N}=2$ level,
\begin{equation}
	(X\overbar X-\nu^2)^5=0~,\label{quintic_constr_phase}
\end{equation}
where $\nu$ is the VEV of $X$ which leads to spontaneous breaking of the $U(1)$. In the limit $\nu=0$, the $U(1)$ is unbroken and the constraint reduces to either $X^3=0$ with complete SUSY breaking, or $X^2=0$ with partial breaking, because the components of the constraints other than these two do not lead to new solutions.

The forms of the two quintic constraints (shift-symmetric and phase-symmetric) are similar to the cubic ${\cal N}=1$ superfield constraints studied in \cite{Aldabergenov:2021obf}, which are given by (calling the chiral superfield $S$)
\begin{align}
	(S+\overbar S)^3 &=0~,\label{N=1_constr_shift}\\
	(S\overbar S-1)^3 &=0~,\label{N=1_constr_phase}
\end{align}
where in \eqref{N=1_constr_phase} we set the VEV of $S$ to one. The constraint \eqref{N=1_constr_phase} can be obtained from \eqref{N=1_constr_shift} by a simple redefinition of the ${\cal N}=1$ chiral superfield, $S\rightarrow\log S$. After the redefinition, \eqref{N=1_constr_shift} becomes $(\log S\overbar S)^3=0$, and we can expand $S\overbar S$ around its unit VEV, $S\overbar S=1+\Gamma$, where $\Gamma$ denotes goldstino-dependent terms:
\begin{equation}
	[\log(1+\Gamma)]^3=\Gamma^3\left(1-\tfrac{1}{2}\Gamma+\ldots\right)^3=0~.
\end{equation}
This is solved by $\Gamma^3=0$, which is exactly the constraint \eqref{N=1_constr_phase}, and implies that the leading component of $\Gamma$ is a (real) bilinear function of the goldstino.

When it comes to ${\cal N}=2$ chiral superfield $X$, the redefinition $X=\log\widetilde X$ cannot be performed. This is because its lowest and highest $\vartheta$-components,
\begin{equation}
	\Phi=\log\widetilde\Phi~,~~~m-\tfrac{1}{4}\overbar D^2\overbar\Phi=\widetilde\Phi^{-1}\left(\widetilde m-\tfrac{1}{4}\overbar D^2\overbar{\widetilde\Phi}-\tfrac{1}{2}\widetilde\Phi^{-1}\widetilde W^2\right)
\end{equation}
are incompatible with each other due to the fact that the "$F$-term" of $X$ is not independent but is a function of $\overbar D^2\overbar\Phi$. Therefore the phase-symmetric ${\cal N}=2$ constraint \eqref{quintic_constr_phase} (and its solution) cannot be derived from $(X+\overbar X)^5=0$ in analogy with the aforementioned ${\cal N}=1$ case, and must be solved separately, which we are going to do next. 

The $\vartheta^2\bar\vartheta^2$-component of \eqref{quintic_constr_phase} reads
\begin{equation}
	\tfrac{1}{8}\Lambda^4(\Box\Lambda+10\widetilde A)+\nu^2\Lambda^3\widetilde A-3\nu^4\Lambda^2\widetilde B+3\nu^4\Lambda W^2\overbar W^2=0~,\label{quintic_subconstr_phase}
\end{equation}
where we introduce for convenience
\begin{equation}
	\Lambda\equiv\Phi\overbar\Phi-\nu^2~,~~~\Sigma\equiv-\frac{i}{2}\log\frac{\Phi}{\overbar\Phi}~,
\end{equation}
and $\widetilde A$ and $\widetilde B$ are defined as
\begin{align}
\begin{split}
\widetilde A &\equiv 2\left(m-\tfrac{1}{4}\overbar D^2\overbar\Phi\right)\left(m-\tfrac{1}{4} D^2\Phi\right)-2\nu^2\partial_m\Sigma\partial^m\Sigma-16W\sigma^m\overbar W\partial_m\Sigma\\
&\hspace{140pt}-\left[iW\sigma^m\partial_m\overbar W-\frac{13}{2\nu}e^{i\Sigma}\left(m-\tfrac{1}{4}\overbar D^2\overbar\Phi\right)\overbar W^2+{\rm h.c.}\right]~,
\end{split}\\
\widetilde B &\equiv-\frac{1}{\nu}e^{i\Sigma}\left(m-\tfrac{1}{4}\overbar D^2\overbar\Phi\right)\overbar W^2-\frac{1}{\nu}e^{-i\Sigma}\left(m-\tfrac{1}{4}D^2\Phi\right)W^2+2W\sigma^m\overbar W\partial_m\Sigma-\frac{4}{\nu^2}W^2\overbar W^2~,
\end{align}

From the lowest component of \eqref{quintic_constr_phase} we have $\Lambda^5=0$. By using the variation of the ${\cal N}=1$ superfields under second supersymmetry,
\begin{equation}
	\delta_{\epsilon_2}\Phi=\sqrt{2}i\epsilon_2^\alpha W_\alpha~,~~~\delta_{\epsilon_2}W_\alpha=-\sqrt{2}i\epsilon_2^\alpha\left(m-\tfrac{1}{4}\overbar D^2\overbar\Phi\right)+\sqrt{2}\sigma^m_{\alpha\dot\alpha}\bar\epsilon_2^{\dot\alpha}\partial_m\Phi~,
\end{equation}
we find that
\begin{equation}
	\Lambda^4W_\alpha=\Lambda^4\overbar W_{\dot\alpha}=0~.\label{eq_Lambda^4_W}\end{equation}
Varying \eqref{eq_Lambda^4_W} shows that $\Lambda^3W^2\propto\Lambda^4$ and so $\Lambda^3W^2\overbar W_{\dot\alpha}=0$. We used these identities in the derivation of Eq. \eqref{quintic_subconstr_phase}.

It is convenient to parametrize the scalar component of $\Phi$ as
\begin{equation}
	\Phi|=|\phi|e^{ia}~,
\end{equation}
where $a$ is our axion in this case, such that $\Sigma|=a$ (the vertical bar denotes $\theta=\bar\theta=0$ component).

Our master equation now is the highest component of Eq. \eqref{quintic_subconstr_phase}, which can be solved (for $\Lambda|$) by the same method we used in the previous subsection. At the leading order in goldstini there are no derivatives of $\Lambda|$ and the solution is (after rescaling $a\rightarrow a/\nu$)
\begin{align}
\begin{split}
\nu^{-1}\Lambda|=& \left[\Omega\overbar\Omega-4(m^2+F\cdot F)\partial a\partial a-8F^{mn}F_{nk}\partial_ma\partial^ka\right]^{-1}\\
&\times\bigg\{\chi^2e^{-ia/\nu}(\bar f\overbar\Omega+2m\,\partial a\partial a)+\lambda^2e^{-ia/\nu}(F\overbar\Omega-2m\,\partial a\partial a)\\
&+\chi\sigma^m\bar\chi\partial^na\left[\eta_{mn}\left(D^2+\tfrac{1}{2}F\cdot F+2|f|^2-2\partial  a\partial a\right)+2D\tilde F_{mn}+2\eta^{kl}F_{mk}F_{ln}\right]\\
&+\lambda\sigma^m\bar\lambda\partial^na\left[\eta_{mn}\left(D^2+\tfrac{1}{2}F\cdot F+2|F|^2-2\partial a\partial a\right)-2D\tilde F_{mn}+2\eta^{kl}F_{mk}F_{ln}\right]\\
&-\sqrt{2}i\overbar\Omega e^{-ia/\nu}(\chi\lambda D+i\chi\sigma^{mn}\lambda F_{mn})\\
&+4\sqrt{2}e^{-ia/\nu}\chi\sigma^{mn}\lambda(F_{mn}\partial a\partial a+2F_{nk}\partial^ka\partial_ma)\\
&-2\sqrt{2}i\chi\sigma^m\bar\lambda\partial^na\left[\overbar F(\eta_{mn}D-iF^+_{mn})-\bar f(\eta_{mn}D-iF^-_{mn})\right]+{\rm h.c.}\bigg\}+\ldots~,
\end{split}
\end{align}
where the ellipsis denotes higher-order terms. Here $\Omega$ is the same as before but with $\phi_2$ replaced by $a$,
\begin{equation}
\Omega\equiv D^2+2\bar fF-\tfrac{1}{2}F\cdot F-\tfrac{i}{2}F\cdot\tilde F-2\partial a\partial a-2i\chi\sigma^m\partial_m\bar\chi-2i\lambda\sigma^m\partial_m\bar\lambda~,
\end{equation}

Since $\Lambda|=|\phi|^2-\nu^2$, we can write $|\phi|$ in terms of $\Lambda|$ as
\begin{equation}
	|\phi|=\nu+\frac{\Lambda|}{2\nu}-\frac{\Lambda|^2}{8\nu^3}+\frac{\Lambda|^3}{16\nu^5}-\frac{5\Lambda|^4}{128\nu^7}~,
\end{equation}
which is an exact expression because $\Lambda^5=0$. Thus the radial scalar $|\phi|$ is eliminated by the quintic ${\cal N}=2$ constraint $(X\overbar X-\nu^2)^5=0$, while the axion $a$ survives along with two goldstini and abelian gauge field.

\subsection{Decoupling ${\cal N}=1$ vector multiplet}\label{subsec_V=0}

Let us go back to the shift-symmetric constraint 
\begin{equation}
	(X+\overbar X)^5=0~,\label{shift_constr_dec}
\end{equation}
and decouple the ${\cal N}=1$ vector multiplet, $\lambda=A_m=D=0$, expecting that the constraint will describe ${\cal N}=1\rightarrow{\cal N}=0$ breaking with a single chiral superfield. In this case the leading component of \eqref{shift_constr_dec} reduces from quintic nilpotent to cubic nilpotent scalar, $(\phi+\bar\phi)^3=0$, because when $\lambda=0$, the saxion $\phi+\bar\phi\equiv 2\phi_1$ becomes a bilinear of $\chi$ and $\bar\chi$, and its third power necessarily vanishes. This also means that in terms of the ${\cal N}=1$ chiral superfield $\Phi$, the constraint becomes,
\begin{equation}
	(\Phi+\overbar\Phi)^3=0~.\label{cubic_N=1}
\end{equation}
It was shown in \cite{Aldabergenov:2021obf} that the cubic constraint \eqref{cubic_N=1} indeed describes spontaneous breaking of ${\cal N}=1$ SUSY, and preserves the axion $\phi_2$, protected by the shift-symmetry, and the goldstino $\chi$.

Let us now reproduce the solution to Eq. \eqref{cubic_N=1} found in \cite{Aldabergenov:2021obf} by simply taking the full solution \eqref{full_sol} to the quintic constraint, and setting $\lambda=A_m=D=0$. Vanishing $\lambda$ means that only the first line in \eqref{full_sol} survives. After some algebra we find
\begin{align}
\begin{split}
\phi_1=& \frac{1}{2U}(\chi^2\overbar F+\bar\chi^2F+2\chi\sigma^m\bar\chi\partial_m\phi_2)\\
&+\left[\frac{i\chi^2}{2U^2}\bar\chi(\overbar\sigma^m\partial_m\chi\overbar F-\partial\bar\chi\partial\phi_2+2\overbar\sigma^{mn}\partial_n\bar\chi\partial_m\phi_2)+{\rm h.c.}\right]\\
&-\frac{\chi^2\bar\chi^2}{2U^3}\Big[\partial_m\bar\chi\overbar\sigma^{mn}\partial_n\bar\chi F+\partial_m\chi\sigma^{mn}\partial_n\chi\overbar F\\
&\hspace{1.6cm}+\partial_n\chi(2\eta^{nk}\sigma^m-\eta^{mk}\sigma^n-\eta^{mn}\sigma^k-i\epsilon^{mnkl}\sigma_l)\partial_k\bar\chi\partial_m\phi_2\Big]~,
\end{split}
\end{align}
where $U\equiv 2(|F|^2-\partial\phi_2\partial\phi_2)$. This solution exactly coincides with the one found in \cite{Aldabergenov:2021obf}. Thus, we conclude that the cubic ${\cal N}=1$ constraint \eqref{cubic_N=1} and its general solution preserving the axion of the $U(1)$ shift symmetry, can be viewed as the special case of the quintic constraint \eqref{shift_constr_dec} where the ${\cal N}=1$ vector multiplet decouples.

\section{Tensor multiplet case}\label{sec_tensor}

The quintic constraints can be applied to another off-shell ${\cal N}=2$ multiplet -- tensor multiplet -- which can be represented by chiral-antichiral superfield, $\overbar D_{\dot\alpha}Y={\cal D}_\alpha Y=0$. It can be expanded as
\begin{equation}
Y=\Phi-\sqrt{2}i\bar{\vartheta}\overbar DL+\bar{\vartheta}^2\left(m-\tfrac{1}{4}\overbar D^2\overbar\Phi\right)~,
\end{equation}
where $\Phi$ is again an ${\cal N}=1$ chiral superfield, and $m$ is a real parameter for partial SUSY breaking. The ${\cal N}=1$ real superfield $L$ is defined by the linearity constraint
\begin{equation}
D^2L=\overbar D^2L=0~,	
\end{equation}
so that its superfield strength $\overbar DL$ is
an ${\cal N}=1$ chiral superfield, and has the component expansion
\begin{equation}
\overbar D_{\dot\alpha}L=\bar\psi_{\dot\alpha}+\theta^\alpha\sigma^m_{\alpha\dot\alpha}(B_m-i\partial_m\varphi)+i\theta^2\partial_m\psi^\alpha\sigma^m_{\alpha\dot\alpha}~,\end{equation}
where $\psi$ is Weyl fermion, $\varphi$ is real scalar, and $B_m\equiv\tfrac{1}{2}\epsilon_{mnkl}\partial^nB^{kl}$ with two-form field $B^{kl}$ encoding a real scalar on-shell degree of freedom (in four dimensions).

We can again impose one of the two quintic constraints
\begin{align}
(Y+\overbar Y)^5 &=0~,\label{Y_constr_shift}\\
(Y\overbar Y-\nu^2) &=0~,\label{Y_constr_phase}
\end{align}
depending on the realization of a global $U(1)$ on $Y$ (or more precisely on its leading component $\Phi$). As in the case of ${\cal N}=2$ vector multiplet, general solutions to the constraints \eqref{Y_constr_shift} and \eqref{Y_constr_phase} eliminate real scalar components $(\Phi+\overbar\Phi)|_{\theta=\bar\theta=0}$ and $|\Phi|_{\theta=\bar\theta=0}$, respectively. The surviving (on-shell) degrees of freedom are then the axion, the real scalar $\varphi$, two goldstini, and the two-form $B_{mn}$. As the linear superfield $L$ does not include any auxiliary fields, in the denominator of the solution we will find the quantity (counterpart of $\Omega$ from Eq. \eqref{Omega_def}),
\begin{equation}
	(\overbar F+m)F+{\rm derivative~terms}~,
\end{equation}
which leads to $\langle F\rangle\neq 0$ (and of course $\langle F\rangle\neq -m$).

\section{A UV model for the quintic constraint}\label{sec_UV}

Here we provide an example of a microscopic theory for ${\cal N}=2$ vector multiplet (within the realm of rigid SUSY), that can lead to the superfield constraint $(X+\overbar X)^5=0$ in the infrared (IR).

In \cite{Dudas:2017sbi} the authors proposed a microscopic theory for the cubic constraint $X^3=0$, which relies on the integral over the whole ${\cal N}=2$ superspace in order to generate large masses for the two scalars, so that they decouple in the IR. Here we follow the same route and introduce the Lagrangian
\begin{equation}
{\cal L}=\left(-\frac{i}{2}\int\!d^2\theta d^2\vartheta{\cal F}(X)+{\rm h.c.}\right)-\gamma\int\!d^4\theta d^4\vartheta{\cal G}(X,\overbar X)~,\label{L_FG}
\end{equation}
where the first term is chiral half-superspace integral of a holomorphic prepotential ${\cal F}(X)$, and leads to the usual two-derivative action for ${\cal N}=2$ vector multiplet with K\"ahler potential, superpotential, and gauge kinetic function given by
\begin{equation}
K=\tfrac{i}{2}(\Phi\overbar{\cal F}_{\bar\Phi}-\overbar\Phi{\cal F}_\Phi)~,~~~{\cal W}=-\tfrac{i}{2}m{\cal F}_\Phi~,~~~g=-i{\cal F}_{\Phi\Phi}~,\label{KWg}
\end{equation}
respectively, where ${\cal F}$ should be understood as a function of $\Phi=X|_{\vartheta=0}$. The second term of \eqref{L_FG} is a real function ${\cal G}(X,\overbar X)$ integrated over the whole ${\cal N}=2$ superspace, which gives rise to higher-derivative terms among other things.

Requiring that the theory is invariant under constant shifts, $X\rightarrow X+i\alpha$, fixes the prepotential as ${\cal F}(X)=\tfrac{i}{2}X^2$, and ${\cal G}$ as ${\cal G}(X+\overbar X)$. Then, performing the $\vartheta$-integration we get the ${\cal N}=1$ superspace formulation of the Lagrangian \eqref{L_FG},
\begin{align}
	&~{\cal L}=\left[\tfrac{1}{2}\int\!d^2\theta\left(-\tfrac{1}{4}\Phi\overbar D^2\overbar\Phi+m\Phi+\tfrac{1}{2}W^2\right)+{\rm h.c.}\right]\nonumber\\
	&-\tfrac{1}{4}\gamma\int\!d^4\theta\bigg\{{\cal G}'\Box\Phi_++4{\cal G}''\left[\left|m-\tfrac{1}{4}\overbar D^2\overbar\Phi\right|^2-\tfrac{1}{4}\partial_m\Phi_-\partial^m\Phi_--\tfrac{i}{2}W\sigma^m\partial_m\overbar W+\tfrac{i}{2}\partial_m W\sigma^m\overbar W\right]\nonumber\\
	&+2{\cal G}'''\left[W^2\left(\overbar m-\tfrac{1}{4}D^2\Phi\right)+\overbar W^2\left(m-\tfrac{1}{4}\overbar D^2\overbar\Phi\right)+iW\sigma^m\overbar W\partial_m\Phi_-\right]+{\cal G}^{(4)}W^2\overbar W^2\bigg\}~,
\end{align}
where $\Phi_\pm\equiv\Phi\pm\overbar\Phi$, ${\cal G}'\equiv\frac{\partial}{\partial\Phi_+}{\cal G}(\Phi_+)$, and ${\cal G}^{(4)}\equiv\frac{\partial^4}{\partial\Phi_+^4}{\cal G}(\Phi_+)$. We discuss the details of this model in Appendix \hyperref[AppB]{B}, while here we focus on the resulting scalar potential. Due to the absence of electric FI term $\xi D$, we have $D=0$ as a solution to its equation of motion. Therefore the potential is determined by the $F$-term part of the Lagrangian,
\begin{equation}
{\cal L}\supset F\overbar F+\tfrac{1}{2}m(F+\overbar F)-\gamma {\cal G}^{(4)}F\overbar F|F+m|^2~,\label{L_of_F}
\end{equation}
and its equation of motion
\begin{equation}
	\overbar F+\tfrac{1}{2}m-2\gamma{\cal G}^{(4)}\overbar F(\overbar F+m)(F+\tfrac{1}{2}m)=0~,\label{F_EOM}
\end{equation}
where ${\cal G}={\cal G}(\phi+\bar\phi)$. Eq. \eqref{F_EOM} has an obvious solution $F=-\tfrac{1}{2}m$, consistent with the limit $\gamma\rightarrow 0$. This gives rise to the scalar potential
\begin{equation}
	{\cal V}=\tfrac{1}{4}m^2+\tfrac{1}{16}\gamma\,m^4{\cal G}^{(4)}~.\label{V_pot_phi}
\end{equation}
As long as $\gamma>0$, a mass term for the real part of $\phi$ can be generated if we take for example
\begin{equation}
	{\cal G}(X+\overbar X)=\frac{1}{1440}(X+\overbar X)^6~,\label{G_mass}
\end{equation}
so that 
\begin{equation}
    {\cal G}^{(4)}(\phi+\bar\phi)=\tfrac{1}{4}(\phi+\bar\phi)^2=\phi_1^2~,
\end{equation}
where $\phi_1$ is the real part of $\phi$.
The imaginary part of $\phi$ -- the axion -- is absent from the potential due to the shift symmetry. We confirm in Appendix \hyperref[AppB]{B} that the choice \eqref{G_mass} does not lead to ghosts and new propagating degrees of freedom.

As can be seen from \eqref{V_pot_phi} and \eqref{G_mass}, the mass of the saxion $\phi_1$ is proportional to $\sqrt{\gamma}m^2$. Thus, at scales much lower than $\sqrt{\gamma}m^2$ (and SUSY breaking scale~\footnote{In this model both supersymmetries are broken at the same scale $\sqrt{m}$ as can be seen from SUSY transformations of the fermions \eqref{lambda_SUSY_transform}.}) the saxion decouples, and the effective theory can be described by the half-superspace integral, 
\begin{equation}
	\tfrac{1}{4}\int\!d^2\theta d^2\vartheta X^2+{\rm h.c.}~,
\end{equation}
with the help of the quintic constraint $(X+\overbar X)^5=0$.

We could also add electric FI terms (with complex parameter $e$ and real $\xi$)
\begin{equation}
	{\cal L}_{\rm FI}=\left(e\!\int\!d^2\theta\,\Phi+{\rm h.c.}\right)+\xi\!\int\!d^4\theta V~,
\end{equation}
to the Lagrangian \eqref{L_FG}, as they are ${\cal N}=2$ supersymmetric. In this case the equations of motion for $F$ and $D$ will have more complicated forms, but it is sufficient that the saxion mass squared, given by $\gamma\langle F\overbar F|F+m|^2\rangle$ (see \eqref{L_of_F}), is non-vanishing and positive.

\section{Conclusion}\label{sec_conclusion}

In this work we studied fifth-order constraint for abelian ${\cal N}=2$ vector multiplet described by (short) chiral-chiral superfield $X$. The constraint takes the form of the nilpotency constraint of degree five on the real superfield $X+\overbar X$. We found most general solution of the constraint, which eliminates the real scalar component of $X$ as a function of the imaginary scalar, two fermions, and the gauge field. This solution necessarily breaks both supersymmetries (thus both Weyl fermions are goldstini) which are non-linearly realized, while the imaginary scalar can be identified as an axion of a broken global $U(1)$ symmetry. Therefore the constraint can be used to describe low-energy dynamics of a microscopic theory of ${\cal N}=2\rightarrow {\cal N}=0$ SUSY breaking by a vector multiplet, where the real scalar component becomes heavy and decouples, while the imaginary scalar is protected by a (exact or approximate) global abelian symmetry. We provide one example of such microscopic theory which is discussed in more detail in Appendix \hyperref[AppB]{B}. 

The constraint can be generalized to the case where an abelian symmetry acts on $X$ (or more precisely on its complex scalar component $\phi$) as a phase rotation, in which case it takes the form,
\begin{equation}
	(X\overbar X-\nu^2)^5=0~,\label{constr_concl}
\end{equation}
where $\nu$ is the VEV of $X$. For $\nu\neq 0$, the $U(1)$ is spontaneously broken and the general solution to \eqref{constr_concl} eliminates the radial scalar $|\phi|$, while preserving the angular part $\log{(\phi/\bar\phi)}$ as the corresponding axion. It is also straightforward to generalize the quintic constraints to ${\cal N}=2$ chiral-antichiral superfield describing ${\cal N}=2$ tensor multiplet.

\begin{figure}
\centering
  \centering
  \includegraphics[width=1\linewidth]{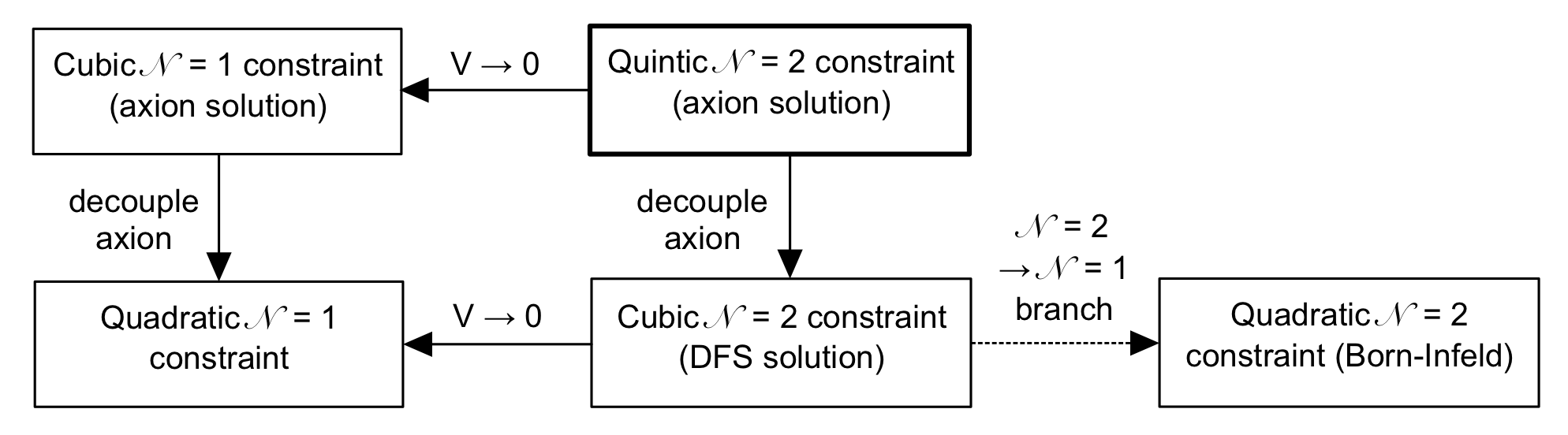}
\captionsetup{width=1\linewidth}
\caption{The relations between various ${\cal N}=2$ and ${\cal N}=1$ superfield constraints and their solutions, stemming from our quintic constraints. $V$ stands for the ${\cal N}=1$ vector multiplet, and DFS for Dudas--Ferrara--Sagnotti \cite{Dudas:2017sbi}. ``Axion solution" is the solution that preserves the corresponding axion.}\label{fig_constr}
\end{figure}

Our quintic superfield constraints seem to be the highest-order -- and thus most general -- constraints that can be imposed on ${\cal N}=2$ vector and tensor multiplets, yielding unique solutions which break both supersymmetries.~\footnote{This is because if we construct a general bilinear function of the goldstini $\chi,\lambda$ and their hermitian conjugates, its fifth power will always vanish.} As such, the quintic constraints and their general solutions reduce in the appropriate limits to various lower-order constraints known in the literature. In particular, in the limit where the ``axion" acquires a large mass and decouples, our quintic constraint, for example $(X+\overbar X)^5=0$, reduces to the cubic chiral constraint $X^3=0$ of Ref. \cite{Dudas:2017sbi}, which eliminates both real scalars. Then, as shown in \cite{Dudas:2017sbi}, one can consistently decouple the ${\cal N}=1$ vector multiplet so that the constraint further reduces to the well-known quadratic constraint for ${\cal N}=1$ chiral superfield, $\Phi^2=0$ \cite{Komargodski:2009rz}. We can also take the following alternative route. First decouple the ${\cal N}=1$ vector multiplet as we described in Subsection \ref{subsec_V=0}: this leads to the ${\cal N}=1$ real cubic constraint $(\Phi+\overbar\Phi)^3=0$ of Ref. \cite{Aldabergenov:2021obf}, which preserves the axion. Then we can eliminate the axion by imposing the stronger quadratic constraint $\Phi^2=0$. If we go back to the ${\cal N}=2$ constraint $X^3=0$, aside from the DFS solution \cite{Dudas:2017sbi}, it also admits a special solution which satisfies $X^2=0$, and describes partial breaking ${\cal N}=2\rightarrow{\cal N}=1$. This solution of course describes supersymmetric Born--Infeld theory \cite{Born:1934gh} with one linear (unbroken) and one non-linear (broken) supersymmetry, where the whole ${\cal N}=1$ chiral superfield decouples, and the ${\cal N}=1$ vector multiplet plays the role of the goldstino multiplet. We summarize the relations between all these constraints in Figure \ref{fig_constr}. The same applies to ${\cal N}=2$ tensor multiplet if we replace ${\cal N}=1$ vector with ${\cal N}=1$ tensor (linear) multiplet.

\section*{Acknowledgements}

This work was partially supported by CUniverse research promotion project of Chulalongkorn University (grant CUAASC), Thailand Science research and Innovation Fund Chulalongkorn University CU$\_$FRB65$\_$ind (2)$\_$107$\_$23$\_$37, and partially performed by I.A. as International professor of the Francqui Foundation, Belgium.

\section*{Appendix A: Full component expression of the quintic constraint}\label{AppA}
\addcontentsline{toc}{section}{\protect\numberline{}Appendix A: Full component expression of the quintic constraint}

In this Appendix we present the full component form of the quintic constraint \eqref{quintic_constr} of Section \ref{sec_quintic_shift}. The component equation is given by \eqref{master_eq}, we repeat it here for convenience,
\begin{align}\label{App_master_eq}
\begin{split}
-\tfrac{2}{3}\phi_1^4\Box^2\phi_1+\phi_1^3\left(P+\tfrac{40}{3}\Box\phi_1\Box\phi_1+\tfrac{16}{3}\partial^{mn}\phi_1\partial_{mn}\phi_1\right)+\phi_1^2\left(J_1+J_2^{mn}\partial_{mn}\phi_1\right)&\\+\phi_1\left(I_1+I_2^{mn}\partial_{mn}\phi_1\right)+H_1+H_2^{mn}\partial_{mn}\phi_1 &=0~.
\end{split}
\end{align}
The expressions for $P,H,I,J$ to all orders in the fermions are,
\begin{align}	
\begin{split}
3P &= 16\Box\phi_2\Box\phi_2+16\partial_{mn}\phi_2\partial^{mn}\phi_2+4\Box(\Omega+i\chi\sigma^m\partial_m\bar\chi+i\lambda\sigma^m\partial_m\bar\lambda)-32\partial_mF\partial^m\overbar F\\
&-16\partial_mD\partial^mD+8\partial^mF_{mn}\partial_kF^{kn}+4\partial_kF_{mn}\partial^kF^{mn}+4i\tilde{F}_{mn}\Box F^{mn}\\
&-16i\Box\chi\sigma^m\partial_m\bar\chi-16i\Box\lambda\sigma^m\partial_m\bar\lambda-16i\partial_{mn}\chi\sigma^m\partial^n\bar\chi-16i\partial_{mn}\lambda\sigma^m\partial^n\bar\lambda+{\rm h.c.}~,
\end{split}
\end{align}
\begin{align}
\begin{split}
	J_1 &= 8i\Omega\Box\phi_2+16\tilde F_{mn}\partial_kF^{nk}\partial^m\phi_2-2\Box(\chi^2\overbar F+\lambda^2f+\chi\sigma^m\bar\chi\partial_m\phi_2+\lambda\sigma^m\bar\lambda\partial_m\phi_2)\\
	&-8i\partial^m(2\bar fF-i\chi\sigma^n\partial_n\bar\chi-i\lambda\sigma^n\partial_n\bar\lambda)\partial_m\phi_2+8(\chi\sigma^m\partial^n\bar\chi+\lambda\sigma^m\partial^n\bar\lambda)\partial_{mn}\phi_2\\
&+8(2\partial^m\chi\sigma_n\partial_m\bar\chi+2\partial^m\lambda\sigma_n\partial_m\bar\lambda-i\epsilon_{mnkl}\partial^m\chi\sigma^k\partial^l\bar\chi-i\epsilon_{mnkl}\partial^m\lambda\sigma^k\partial^l\bar\lambda)\partial^n\phi_2\\
&-8\partial_m(\chi\sigma^m\overbar\sigma^n\partial_n\chi\bar f)-8\partial_m(\lambda\sigma^m\overbar\sigma^n\partial_n\lambda F)-4\sqrt{2}i\partial^m\chi\partial^n\lambda(\eta_{mn}D+2iF^+_{mn})\\
&+16\sqrt{2}i\partial_m\chi\sigma^{mn}\partial_n\lambda D-4\sqrt{2}(\partial^m\chi\sigma^{nk}\partial_k\lambda-\partial^m\lambda\sigma^{nk}\partial_k\chi)(F^+_{mn}+2i\tilde F_{mn})\\
&-2\sqrt{2}\chi\sigma^{mn}\Box\lambda F_{mn}+2\sqrt{2}\lambda\sigma^{mn}\Box\chi F_{mn}-8\sqrt{2}\chi\sigma^{mn}\partial_n\partial^k\lambda F_{mk}\\
&+8\sqrt{2}\lambda\sigma^{mn}\partial_n\partial^k\chi F_{mk}-16\sqrt{2}\chi\sigma^{mn}\partial_m\lambda\partial^k F_{nk}+16\sqrt{2}\lambda\sigma^{mn}\partial_m\chi\partial^k F_{nk}\\
&-8\sqrt{2}i(\chi\sigma^{mn}\partial_m\lambda+\lambda\sigma^{mn}\partial_m\chi)\partial_n D-16\chi\Box\chi\bar f-16\lambda\Box\lambda F\\
&+6\sqrt{2}i(\chi\Box\lambda+\lambda\Box\chi)D-4\sqrt{2}(\chi\partial^m\lambda-\lambda\partial^m\chi)\partial^nF_{mn}+8\partial\chi^2\partial\overbar F\\
&+8\partial\lambda^2\partial F-8\sqrt{2}i\partial_m(\chi\lambda)\partial^mD+2\sqrt{2}i\chi\lambda\Box D-2\sqrt{2}\chi\sigma^{mn}\lambda\Box F_{mn}+{\rm h.c.}~,
\end{split}
\end{align}
\begin{align}
\begin{split}
J_2^{mn} &= 4\eta^{mn}(3\Omega+4\partial\phi_2\partial\phi_2+m^2+F\cdot F+i\chi\sigma^k\partial_k\bar\chi+i\lambda\sigma^k\partial_k\bar\lambda)+8\eta_{kl}F^{mk}F^{ln}\\
&-16\partial^m\phi_2\partial^n\phi_2-8i\chi\sigma^m\partial^n\bar\chi-8i\lambda\sigma^m\partial^n\bar\lambda+{\rm h.c.}~,
\end{split}
\end{align}
\begin{align}
\begin{split}
	I_1 &= 2\Omega\overbar\Omega-8(m^2+F\cdot F)\partial\phi_2\partial\phi_2-16F^{mn}F_{nk}\partial^k\phi_2\partial_m\phi_2\\
	&-8i\chi\sigma^m\partial_m\bar\chi(|f|^2+\partial\phi_2\partial\phi_2)-8i\lambda\sigma^m\partial_m\bar\lambda(|F|^2+\partial\phi_2\partial\phi_2)\\
	&-4i(\chi\sigma^m\partial^n\bar\chi+\lambda\sigma^m\partial^n\bar\lambda)\left[\eta_{mn}(D^2+\tfrac{1}{2}F\cdot F)+2\eta^{kl}F_{nl}F_{km}-4\partial_m\phi_2\partial_n\phi_2\right]\\
	&-8i(\chi\sigma^m\partial^n\bar\chi-\lambda\sigma^m\partial^n\bar\lambda)D\tilde F_{mn}-8i(\chi^2\bar f+\lambda^2 F-\sqrt{2}i\chi\lambda D+\sqrt{2}\chi\sigma^{mn}\lambda F_{mn})\Box\phi_2\\
	&-4\sqrt{2}\overbar F(\chi\sigma^m\partial^n\bar\lambda-\partial^n\chi\sigma^m\bar\lambda-\chi\sigma^m\bar\lambda\partial^n)(\eta_{mn}D-iF^+_{mn})\\
	&-4\sqrt{2}f(\lambda\sigma^m\partial^n\bar\chi-\partial^n\lambda\sigma^m\bar\chi-\lambda\sigma^m\bar\chi\partial^n)(\eta_{mn}D+iF^+_{mn})\\
	&+8i\big(m\,\partial^m\chi^2-m\,\partial^m\lambda^2 -4\chi\sigma^{mn}\partial_n\chi\bar f-4\lambda\sigma^{mn}\partial_n\lambda F\\
	&\hspace{6cm}+2\sqrt{2}i\chi\sigma^{mn}\partial_n\lambda D+2\sqrt{2}i\lambda\sigma^{mn}\partial_n\chi D\big)\partial_m\phi_2\\
	&-8\sqrt{2}i(\chi\partial^m\lambda-\lambda\partial^m\chi)\partial^n\phi_2F^+_{mn}+8\sqrt{2}i(\chi\sigma^{mn}\partial^k\lambda-\lambda\sigma^{mn}\partial^k\chi)\partial_m\phi_2 F^-_{nk}\\
	&+8\sqrt{2}i(\chi\sigma^{mn}\partial_n\lambda-\lambda\sigma^{mn}\partial_n\chi)\partial^k\phi_2(F^+_{mk}+2i\tilde F_{mk})\\
	&+\chi^2(8i\partial\overbar F\partial\phi_2+\Box\bar\chi^2+4\lambda\Box\lambda)+\lambda^2(8i\partial F\partial\phi_2+\Box\bar\lambda^2+4\chi\Box\chi)\\
	&+4\chi\sigma^m\bar\chi(2if\partial_m\overbar F+D\partial^nF_{mn}-\partial^nDF_{mn}+\tilde{F}_{mk}\partial_nF^{kn}+\partial_n\lambda\sigma^n\partial_m\bar\lambda-\lambda\sigma^n\partial_{mn}\bar\lambda)\\
	&+4\lambda\sigma^m\bar\lambda(2i\overbar F\partial_m F-D\partial^nF_{mn}+\partial^nDF_{mn}+\tilde{F}_{mk}\partial_nF^{kn}+\partial_n\chi\sigma^n\partial_m\bar\chi-\chi\sigma^n\partial_{mn}\bar\chi)\\
	&+8\sqrt{2}\chi\lambda\partial\phi_2\partial D+16\sqrt{2}i\chi\sigma^{mn}\lambda\partial_n\phi_2\partial^kF_{mk}+4\partial_m(\chi^2\lambda)\sigma^m\overbar\sigma^n\partial_n\lambda\\
	&+4\partial_m(\lambda^2\chi)\sigma^m\overbar\sigma^n\partial_n\chi-8\sqrt{2}\chi\sigma^m\bar\lambda\partial^n\overbar F(\eta_{mn}D-iF_{mn})-\partial\chi^2\partial\bar\chi^2-\partial\lambda^2\partial\bar\lambda^2\\
	&-2(\chi\sigma^m\partial^n\bar\chi-\partial^n\chi\sigma^m\bar\chi)(\lambda\sigma_n\partial_m\bar\lambda-\partial_m\lambda\sigma_n\bar\lambda)\\
	&-2(\chi\sigma^m\partial_m\bar\chi-\partial_m\chi\sigma^m\bar\chi)(\lambda\sigma^n\partial_n\bar\lambda-\partial_n\lambda\sigma^n\bar\lambda)+{\rm h.c.}~,
\end{split}
\end{align}
\begin{align}
\begin{split}
I_2^{mn} &= -4\eta^{mn}\Big[\chi^2(\overbar F+2\bar f)+\lambda^2(f+2F)+(\chi\sigma^k\bar\chi+\lambda\sigma^k\bar\lambda)\partial_k\phi_2\\&\hspace{8cm}-3\sqrt{2}i\chi\lambda D+\sqrt{2}\chi\sigma^{kl}\lambda F_{kl}\Big]\\
&-8(\chi\sigma^m\bar\chi+\lambda\sigma^m\bar\lambda)\partial^n\phi_2+16\sqrt{2}\chi\sigma^{ml}\lambda F_{lk}\eta^{kn}+{\rm h.c.}~,
\end{split}
\end{align}
\begin{align}
\begin{split}
	H_1 &= -\chi^2\big[2\bar f\overbar\Omega+4m\partial\phi_2\partial\phi_2-2i\overbar F(\lambda\sigma^m\partial_m\bar\lambda-\partial_m\lambda\sigma^m\bar\lambda)+2i\partial\bar\chi^2\partial\phi_2\\
	&\qquad-4i\lambda\sigma^m\overbar\sigma^n\partial_n\lambda\partial_m\phi_2-2i\lambda\sigma^m\bar\lambda\partial_m\overbar F+\sqrt{2}\lambda\sigma^m\partial_m(\bar\chi D)+\sqrt{2}i\lambda\sigma^m\partial^n(\bar\chi F^-_{mn})\big]\\
	&-\lambda^2\big[2F\overbar\Omega-4 m\partial\phi_2\partial\phi_2-2if(\chi\sigma^m\partial_m\bar\chi-\partial_m\chi\sigma^m\bar\chi)+2i\partial\bar\lambda^2\partial\phi_2\\
	&\qquad-4i\chi\sigma^m\overbar\sigma^n\partial_n\chi\partial_m\phi_2-2i\chi\sigma^m\bar\chi\partial_m F+\sqrt{2}\chi\sigma^m\partial_m(\bar\lambda D)-\sqrt{2}i\chi\sigma^m\partial^n(\bar\lambda F^-_{mn})\big]\\
	&-\chi\sigma^m\bar\chi\partial_m\phi_2(2D^2+F\cdot F+4|f|^2-4\partial\phi_2\partial\phi_2-4i\lambda\sigma^n\partial_n\bar\lambda)\\
	&-4\chi\sigma^m\bar\chi\partial^n\phi_2(D\tilde F_{mn}+\eta^{kl}F_{ml}F_{kn}-i\lambda\sigma_n\partial_m\bar\lambda)\\
	&-\lambda\sigma^m\bar\lambda\partial_m\phi_2(2D^2+F\cdot F+4|F|^2-4\partial\phi_2\partial\phi_2-4i\chi\sigma^n\partial_n\bar\chi)\\
	&-4\lambda\sigma^m\bar\lambda\partial^n\phi_2(-D\tilde F_{mn}+\eta^{kl}F_{ml}F_{kn}-i\chi\sigma_n\partial_m\bar\chi)\\
	&+2\sqrt{2}i\overbar\Omega(\chi\lambda D+i\chi\sigma^{mn}\lambda F_{mn})-8\sqrt{2}\chi\sigma^{mn}\lambda(F_{mn}\partial\phi_2\partial\phi_2+2F_{nk}\partial^k\phi_2\partial_m\phi_2)\\
	&+4\sqrt{2}i\chi\sigma^m\bar\lambda\partial^n\phi_2\left[\overbar F(\eta_{mn}D-iF^+_{mn})-\bar f(\eta_{mn}D-iF^-_{mn})\right]\\
	&+\sqrt{2}\chi\sigma^m\partial^n(\bar\lambda\bar\chi^2)(\eta_{mn}D-iF^+_{mn})+\sqrt{2}\lambda\sigma^m\partial^n(\bar\chi\bar\lambda^2)(\eta_{mn}D+iF^+_{mn})\\
	&-2i\chi\sigma^m\bar\chi\partial_m\lambda^2f-2i\lambda\sigma^m\bar\lambda\partial_m\chi^2\overbar F+2i\chi^2\lambda^2\Box\phi_2+{\rm h.c.}~,
\end{split}\\[10pt]
	H_2^{mn} &= \eta^{mn}(\chi^2\bar\chi^2+\lambda^2\bar\lambda^2+2\chi^2\lambda^2+2\bar\chi^2\bar\lambda^2)-4\chi\sigma^m\bar\chi\lambda\sigma^n\bar\lambda~.
\end{align}
$\Omega$ is defined in \eqref{Omega_def}.

\section*{Appendix B: Details of the UV model}\label{AppB}
\addcontentsline{toc}{section}{\protect\numberline{}Appendix B: Details of the UV model}

The Lagrangian of the UV model of Section \ref{sec_UV} in ${\cal N}=2$ superspace reads
\begin{equation}
{\cal L}=\left(-\frac{i}{2}\int\!d^2\theta d^2\vartheta{\cal F}(X)+{\rm h.c.}\right)-\gamma\int\!d^4\theta d^4\vartheta{\cal G}(X,\overbar X)~,\label{L_N=2}
\end{equation}
where ${\cal F}=\tfrac{i}{2}X^2$, and ${\cal G}$ is a function of $X+\overbar X$, as required by the global symmetry $X\rightarrow X+i\alpha$.

After integrating over $\vartheta$, we obtain the ${\cal N}=1$ form of the Lagrangian for which we introduce the following notation. The first term in \eqref{L_N=2} is denoted ${\cal L}_1$, and is given by
\begin{equation}
    {\cal L}_1=\tfrac{1}{2}\int\!d^2\theta\left(-\tfrac{1}{4}\Phi\overbar{D}^2\overbar\Phi+m\Phi+\tfrac{1}{2}W^2\right)+{\rm h.c.}~.
\end{equation}
For convenience we divide the second term of \eqref{L_N=2} into seven parts (integrated over $\vartheta,\bar\vartheta$):
\begin{gather}
\begin{gathered}
    {\cal L}_2=-\tfrac{\gamma}{4}\int\!d^4\theta{\cal G}'\Box\Phi_+~,~~~{\cal L}_3=-\gamma\int\!d^4\theta{\cal G}''(m-\tfrac{1}{4}D^2\Phi)(m-\tfrac{1}{4}\overbar{D}^2\overbar\Phi)~,\\
    {\cal L}_4=\tfrac{\gamma}{4}\int\!d^4\theta{\cal G}''\partial_m\Phi_-\partial^m\Phi_-~,~~~{\cal L}_5=\tfrac{\gamma i}{2}\int\!d^4\theta{\cal G}''(W\sigma^m\partial_m\overbar W-{\rm h.c.})~,\\
    {\cal L}_6=-\tfrac{\gamma}{2}\int\!d^4\theta{\cal G}'''W^2(m-\tfrac{1}{4}D^2\Phi)+{\rm h.c.}~,~~~{\cal L}_7=-\tfrac{\gamma i}{2}\int\!d^4\theta{\cal G}'''W\sigma^m\overbar W\partial_m\Phi_-~,\\
    {\cal L}_8=-\tfrac{\gamma}{4}\int\!d^4\theta{\cal G}^{(4)}W^2\overbar W^2~.
\end{gathered}
\end{gather}
Next, let us find the full component Lagrangian up to two fermions, using the expansion of $\Phi$ and $W$ given by \eqref{W_Phi_exp}.
${\cal L}_1$ is the usual (two-derivative) action for the ${\cal N}=2$ vector multiplet,
\begin{align}
\begin{split}
    {\cal L}_1=-\partial\phi_1\partial\phi_1-\partial\phi_2\partial\phi_2 -i\chi\sigma^m\partial_m\bar\chi &-i\lambda\sigma^m\partial_m\bar\lambda-\tfrac{1}{4}F_{mn}F^{mn}\\
    &+\tfrac{1}{2}m(F+\overbar F)+F\overbar F+\tfrac{1}{2}D^2~,
\end{split}
\end{align}
while ${\cal L}_2,...,{\cal L}_8$ extend the simplest model by higher-order interactions and higher derivatives. They are given by (up to total derivatives)
\begin{align}
    \begin{split}
        \tfrac{4}{\gamma}{\cal L}_2 &=-{\cal G}'''\Box\phi_1(F\overbar F-\partial\phi_2\partial\phi_2-i\chi\sigma^m\partial_m\bar\chi)+{\cal G}^{(4)}\Box\phi_1(\chi^2\overbar F+\chi\sigma^m\bar\chi\partial_m\phi_2)\\
        &+{\cal G}'''(\chi\sigma^m\Box\bar\chi\partial_m\phi_2+\tfrac{1}{2}\chi\sigma^m\bar\chi\partial_m\Box\phi_2+\chi\Box\chi\overbar F+\tfrac{1}{2}\chi^2\Box\overbar F)\\
        &-{\cal G}''(\tfrac{1}{2}\Box\phi_1\Box\phi_1+\tfrac{i}{2}\partial_m\chi\sigma^m\Box\bar\chi-\tfrac{i}{2}\chi\sigma^m\partial_m\Box\bar\chi+\overbar F\Box F-\partial_m\phi_2\partial^m\Box\phi_2)\\
        &-\tfrac{1}{4}{\cal G}'\Box^2\phi_1+{\rm h.c.}~,
    \end{split}\\[10pt]
    \begin{split}
        \tfrac{1}{\gamma}{\cal L}_3 &=\tfrac{1}{2}{\cal G}^{(5)}f\bar f(\chi^2\overbar F+\chi\sigma^m\bar\chi\partial_m\phi_2)-\tfrac{1}{2}{\cal G}^{(4)}f\bar f(F\overbar F-\partial\phi_2\partial\phi_2-i\chi\sigma^m\partial_m\bar\chi)\\
        &+\tfrac{1}{4}{\cal G}^{(4)}\big[2if\overbar F\chi\sigma^m\partial_m\bar\chi+\chi^2\bar f\Box(\phi_1+i\phi_2)+i\chi\sigma^m\bar\chi\bar f\partial_mF+2i\bar f\chi\sigma^n\overbar\sigma^m\partial_m\chi\partial_n\phi_2\big]\\
        &-\tfrac{1}{2}{\cal G}'''\big[2f\overbar F\Box(\phi_1-i\phi_2)-2i\bar f\partial_m\phi_2\partial^m F+\tfrac{1}{2}f\bar f\Box\phi_1-\bar f\chi\Box\chi-\chi\sigma^m\overbar\sigma^n\partial_n\chi\partial_m\overbar F\\
        &\hspace{2cm}-2i\chi\sigma^m\partial_m\bar\chi\Box(\phi_1+i\phi_2)-\bar f\partial_m\chi\sigma^m\overbar\sigma^n\partial_n\chi+\partial_m\chi\sigma^m\overbar\sigma^n\sigma^k\partial_k\bar\chi\partial_n\phi_2\big]\\
        &-\tfrac{1}{2}{\cal G}''(\Box\phi_1\Box\phi_1+\Box\phi_2\Box\phi_2+\tfrac{1}{2}\bar f\Box F-\tfrac{1}{2}\partial F\partial\overbar F-i\Box\chi\sigma^m\partial_m\bar\chi)+{\rm h.c.}~,
    \end{split}\\[10pt]
    \begin{split}
        \tfrac{2}{\gamma}{\cal L}_4 &={\cal G}^{(5)}\partial\phi_2\partial\phi_2(\chi^2\overbar F+\chi\sigma^m\bar\chi\partial_m\phi_2)-{\cal G}^{(4)}\partial\phi_2\partial\phi_2(F\overbar F-\partial\phi_2\partial\phi_2-i\chi\sigma^m\partial_m\bar\chi)\\
        &+{\cal G}^{(4)}(2i\chi\sigma^m\partial^n\bar\chi\partial_m\phi_2\partial_n\phi_2-2i\chi\partial_m\chi\partial^m\phi_2\overbar F-\chi\sigma^m\bar\chi\partial^n\phi_2\partial_{mn}\phi_1+i\chi^2\partial_m\phi_2\partial^m\overbar F)\\
        &-{\cal G}'''\big(\tfrac{1}{2}\partial\phi_2\partial\phi_2\Box\phi_1+2\partial^m\phi_2\partial^n\phi_2\partial_{mn}\phi_1-2i\overbar F\partial_m F\partial^m\phi_2+\tfrac{1}{2}\partial\chi\partial\chi\overbar F\\
        &\hspace{5cm}-\chi\partial_m\chi\partial^m\overbar F-\partial_n\chi\sigma^n\partial^m\bar\chi\partial_m\phi_2-\tfrac{1}{2}\partial^m\chi\sigma^n\partial_m\bar\chi\partial_n\phi_2\\
        &\hspace{8cm}+i\chi\sigma^n\partial^m\bar\chi\partial_{mn}\phi_1+\chi\sigma^n\partial_{mn}\bar\chi\partial^m\phi_2\big)\\
        &-\tfrac{1}{2}{\cal G}''(\partial^m\phi_2\partial_m\Box\phi_2+\partial F\partial\overbar F-\partial_{mn}\phi_1\partial^{mn}\phi_1-\tfrac{i}{2}\partial^m\chi\sigma^n\partial_{mn}\bar\chi)+{\rm h.c.}~,
    \end{split}
\end{align}
\begin{align}
    \begin{split}
        \tfrac{4}{\gamma}{\cal L}_5 &={\cal G}''\partial_k(\eta_{mn}D-iF^+_{mn})\partial^n(\eta^{km}D+iF_-^{km})-{\cal G}''(\eta^{mk}D-iF^{mk}_+)\partial_k\partial^n(\eta_{mn}D-iF^-_{mn})\\
        &+2i{\cal G}'''\partial_k\phi_2(\eta_{mn}D-iF^+_{mn})\partial^n(\eta^{km}D+iF_-^{km})+2i{\cal G}^{(4)}(F\overbar F-\partial\phi_2\partial\phi_2)\lambda\sigma^m\partial_m\bar\lambda\\
        &+\sqrt{2}{\cal G}^{(4)}(\chi\sigma^k\overbar\sigma^m\partial^n\lambda\partial_k\phi_2+\tfrac{1}{2}\chi\sigma^k\overbar\sigma^m\lambda\partial_k\phi_2\partial^n-\lambda\sigma^m\bar\chi F\partial^n)(\eta_{mn}D+iF^-_{mn})\\
        &+i{\cal G}^{(4)}\chi\sigma_k\bar\chi(\eta_{mn}D-iF^+_{mn})\partial^n(\eta^{mk}D-iF_-^{mk})\\
        &+\sqrt{2}{\cal G}^{(4)}\chi\sigma^m\partial^n\bar\lambda\overbar F(\eta_{mn}D-iF^+_{mn})-\sqrt{2}i{\cal G}'''(\chi\Box\lambda D+i\chi\sigma^{mn}\Box\lambda F_{mn})\\
        &+\tfrac{i}{2}{\cal G}''(5\Box\lambda\sigma^m\partial_m\bar\lambda+\lambda\sigma^m\partial_m\Box\bar\lambda-2i\partial^n\lambda\sigma^m\partial_{mn}\bar\lambda)\\
        &+2{\cal G}'''\big(\lambda\sigma^m\partial_{mn}\bar\lambda\partial^n\phi_2-\partial_n\lambda\sigma^m\partial_m\bar\lambda\partial^n\phi_2+\tfrac{i}{2}\lambda\sigma^m\partial_m\bar\lambda\Box\phi_1\\
        &\hspace{9cm}-\partial_m\lambda\sigma^m\overbar\sigma^n\partial_n\lambda F+\lambda\Box\lambda F\big)\\
        &+\sqrt{2}i{\cal G}'''\big[\chi\sigma^m\overbar\sigma^k(\partial_k\lambda+\lambda\partial_k)\partial^n(\eta_{mn}D-iF^+_{mn})-\tfrac{1}{4}\partial_k\chi\sigma^k\overbar\sigma^m\lambda\partial^n(\eta_{mn}D+iF^-_{mn})\big]\\
        &-\tfrac{\sqrt{2}i}{4}{\cal G}'''\big[\chi\sigma^k\overbar\sigma^m(\partial_k\lambda+\lambda\partial_k)\partial^n+2\chi\sigma^k\overbar\sigma^m(\partial_k\partial^n\lambda-\partial^n\lambda\partial_k)\big](\eta_{mn}D+iF^-_{mn})\\
        &-\tfrac{i}{\sqrt{2}}{\cal G}'''\partial_k\chi\sigma^k\overbar\sigma^m\partial^n\lambda(\eta_{mn}D+iF^-_{mn})+{\rm h.c.}~,
    \end{split}\\[10pt]
    \begin{split}
        \tfrac{2}{\gamma}{\cal L}_6 &=-{\cal G}^{(4)}f\overbar F(D^2-\tfrac{1}{2}F\cdot F-\tfrac{i}{2}F\cdot\tilde F)-{\cal G}'''\Box\phi_1(D^2-\tfrac{1}{2}F\cdot F)+\tfrac{1}{2}{\cal G}'''\Box\phi_2F\cdot\tilde F\\
        &+{\cal G}^{(5)}\lambda^2f(F\overbar F-\partial\phi_2\partial\phi_2)+\tfrac{1}{2}{\cal G}^{(5)}\chi^2\bar f(D^2-\tfrac{1}{2}F\cdot F+\tfrac{i}{2}F\cdot\tilde F)\\
        &-\sqrt{2}i{\cal G}^{(5)}f\big[\lambda\sigma^m\bar\chi\partial^n\phi_2(\eta_{mn}D+iF^+_{mn})+\overbar F(\chi\lambda D+i\chi\sigma^{mn}\lambda F_{mn})\big]\\
        &+i{\cal G}^{(4)}\big[\chi\sigma^m\partial_m\bar\chi(D^2-\tfrac{1}{2}F\cdot F+\tfrac{i}{2}F\cdot\tilde F)+2f(\lambda\sigma^m\partial_m\bar\lambda\overbar F+\lambda\partial\lambda\partial\phi_2-\tfrac{i}{4}\lambda^2\Box\phi_1)\big]\\
        &-\tfrac{1}{\sqrt{2}}{\cal G}^{(4)}\big[\partial^n\lambda\sigma^m\bar\chi f-\lambda\sigma^m\partial^n\bar\chi f-2i\lambda\sigma^m\overbar\sigma^k\partial_k\chi\partial^n(\phi_1-i\phi_2)\big](\eta_{mn}+iF^+_{mn})\\
        &+{\cal G}^{(4)}\lambda^2(\Box\phi_1+i\Box\phi_2-i\partial_m\phi_2\partial^m)F+\tfrac{1}{2}{\cal G}'''(\Box\lambda^2+\tfrac{1}{2}\lambda^2\Box-\partial_m\lambda^2\partial^m)f\\
        &+2i{\cal G}'''\lambda\sigma^m\partial_m\bar\lambda\Box(\phi_1+i\phi_2)-\sqrt{2}i{\cal G}^{(4)}(\chi\lambda D-i\chi\sigma^{mn}\lambda F_{mn})\Box(\phi_1+i\phi_2)\\
        &-\sqrt{2}i{\cal G}'''(\lambda\Box\chi D-i\lambda\sigma^{mn}\Box\chi F_{mn})+\tfrac{1}{\sqrt{2}}{\cal G}^{(4)}\lambda\sigma^m\bar\chi(\partial^n F-f\partial^n)(\eta_{mn}D+iF^+_{mn})
    \end{split}\\[10pt]
    \begin{split}
        \tfrac{2}{\gamma}{\cal L}_7 &=({\cal G}^{(4)}\partial^n\phi_2\partial_k\phi_2-\tfrac{1}{2}{\cal G}'''\partial^n\partial_k\phi_1)(\delta^k_nD^2+\tfrac{1}{2}\delta^k_nF\cdot F+2F^{km}F_{mn})\\
        &-{\cal G}'''\partial^n\phi_2(2D\partial^mF_{mn}+\partial_kF^{km}\tilde F_{mn})-2{\cal G}^{(4)}\partial_m\phi_1\partial_n\phi_2F^{mn}D\\
        &-{\cal G}^{(5)}\lambda\sigma^m\bar\lambda\partial_m\phi_2\partial\phi_2\partial\phi_2+{\cal G}^{(4)}\lambda\sigma^m\bar\lambda(\partial^n\phi_2\partial_{mn}\phi_1+\tfrac{1}{2}\partial_m\phi_2\Box\phi_1-i\overbar F\partial_m F)\\
        &+{\cal G}'''\big[2\partial_m\lambda\sigma^m\partial^n\bar\lambda\partial_n\phi_2+i\lambda\sigma^m\partial^n\bar\lambda\partial_{mn}(\phi_1-i\phi_2)+\tfrac{1}{4}\lambda\sigma^m\bar\lambda\partial_m\Box\phi_2\\
        &\quad+\tfrac{1}{2}\Box\lambda\sigma^m\bar\lambda\partial_m\phi_2-\tfrac{3}{2}\partial_n\lambda\sigma^m\partial^n\bar\lambda\partial_m\phi_2+i\epsilon^{mnkl}\partial_n\lambda\sigma_l\partial_k\bar\lambda\partial_m\phi_2+\lambda\sigma^m\overbar\sigma^n\partial_n\lambda\partial_m F\big]\\
        &-{\cal G}^{(4)}\lambda\sigma^m\big[2i(\partial^n\bar\lambda\partial_n\phi_2-\overbar\sigma^n\partial_n\lambda F)\partial_m\phi_2+\tfrac{1}{\sqrt{2}}(\partial^n\bar\chi F-\bar\chi\partial^n F)(\eta_{mn}D+iF^-_{mn})\big]\\
        &+\tfrac{1}{2}\chi\sigma_k(i{\cal G}^{(4)}\partial^n\bar\chi+{\cal G}^{(5)}\bar\chi\partial^n\phi_2)(\delta^k_nD^2+\tfrac{1}{2}\delta^k_nF\cdot F+2F^{km}F_{mn}+2\eta^{km}D\tilde F_{mn})\\
        &+\sqrt{2}i{\cal G}^{(5)}\lambda\sigma^m(\bar\chi F-\overbar\sigma^k\chi\partial_k\phi_2)\partial^n\phi_2(\eta_{mn}D+iF^-_{mn})\\
        &-\tfrac{1}{\sqrt{2}}{\cal G}^{(4)}\lambda\sigma^m\overbar\sigma^k(\partial_k\chi\partial^n\phi_2+\partial^n\chi\partial_k\phi_2)(\eta_{mn}D+iF^-_{mn})\\
        &+\sqrt{2}\partial_k\lambda\sigma^k\overbar\sigma^m({\cal G}^{(4)}\chi\partial^n\phi_2-\tfrac{i}{2}{\cal G}'''\partial^n\chi)(\eta_{mn}D-iF^+_{mn})\\
        &-\tfrac{1}{\sqrt{2}}{\cal G}^{(4)}\partial^n\phi_2\chi\sigma^k\overbar\sigma^m(\partial_k\lambda-\lambda\partial_k)(\eta_{mn}D+iF^-_{mn})\\
        &+\tfrac{i}{\sqrt{2}}\lambda\sigma^m\overbar\sigma^k({\cal G}^{(4)}\chi\partial^n\partial_k\phi_1-{\cal G}^{(4)}\partial^n\chi\partial_k\phi_1-{\cal G}'''\partial^n\chi\partial_k)(\eta_{mn}D+iF^-_{mn})+{\rm h.c.}~,
    \end{split}
\end{align}
\begin{align}
    \begin{split}
        \tfrac{8}{\gamma}{\cal L}_8 &=-{\cal G}^{(4)}|D^2-\tfrac{1}{2}F\cdot F+\tfrac{i}{2}F\cdot\tilde F|^2+2i{\cal G}^{(4)}\lambda\sigma^m\partial_m\bar\lambda(3D^2-\tfrac{1}{2}F\cdot F+iF\cdot\tilde F)\\
        &-4i{\cal G}^{(4)}\lambda\sigma_k\partial^n\bar\lambda(F_{mn}F^{mk}-\eta_{mn}\tilde F^{mk}D)+{\cal G}^{(5)}\lambda^2F(2D^2-F\cdot F+iF\cdot\tilde F)\\
        &+2{\cal G}^{(5)}\lambda\sigma_k\bar\lambda\partial^n\phi_2(\delta^k_nD^2+\tfrac{1}{2}\delta^k_nF\cdot F-2F_{mn}F^{mk}+2\eta_{mn}\tilde F^{mk}D)\\
        &-2{\cal G}^{(4)}\lambda\sigma_m\bar\lambda(F^{mn}\partial_nD-D\partial_nF^{mn}+\tilde F^{mn}\partial^kF_{nk})-4{\cal G}^{(4)}(\lambda\sigma^m\partial_m\bar\lambda)(\partial_n\lambda\sigma^n\bar\lambda)\\
        &+{\cal G}^{(4)}\partial\lambda^2\partial\bar\lambda^2-\sqrt{2}i{\cal G}^{(5)}(\chi\lambda D+i\chi\sigma^{mn}\lambda F_{mn})(2D^2-F\cdot F+iF\cdot\tilde F)+{\rm h.c.}~,
    \end{split}
\end{align}
where we ignored terms with three or more fermions, and denoted $f\equiv m+F$ as before (also $\Phi_\pm\equiv\Phi\pm\overbar\Phi$, and same for its scalar component $\phi$).

As long as the derivatives of ${\cal G}$ w.r.t. $\phi_+$ (up to fifth) vanish at the vacuum, new degrees of freedom are not generated, since all the terms from ${\cal L}_{2,...,8}$ will vanish as well. For example the possible kinetic terms $\partial F\partial\overbar F$ and $\partial D\partial D$ multiply ${\cal G}''$, so we require $\langle{\cal G}''\rangle=0$ in order to keep $F$ and $D$ fields auxiliary. Another question is whether or not the $\gamma$-terms can introduce ghost instabilities for the existing propagating fields $\phi_1,\phi_2,\chi,\lambda,A_m$. To answer this, let us write down the terms (from the entire Lagrangian) containing at most two derivatives in the bosonic sector, and at most one derivative in the fermionic sector. Then for the bosonic Lagrangian we have
\begin{align}
\begin{split}
    {\cal L}_{\rm bos}= &-\big[1-\gamma{\cal G}^{(4)}(m^2+3mF+3m\overbar F+6F\overbar F+3D^2)\big]\partial\phi_1\partial\phi_1\\
    &-\big[1-\gamma{\cal G}^{(4)}(m^2+mF+m\overbar F+D^2)\big]\partial\phi_2\partial\phi_2\\
    &-\tfrac{1}{4}\big[1-\gamma{\cal G}^{(4)}(mF+m\overbar F+2F\overbar F+D^2)\big]F_{mn}F^{mn}\\
    &+\tfrac{1}{2}m(F+\overbar F)+F\overbar F+\tfrac{1}{2}D^2-\gamma{\cal G}^{(4)}|mF+F\overbar F+\tfrac{1}{2}D^2|^2+\ldots~,\label{L_bos_F}
\end{split}
\end{align}
where $\ldots$ stands for irrelevant terms containing derivatives of the auxiliary fields and derivative interactions between $\phi_1$ and $\phi_2$ (since these do not contribute to the kinetic terms).

It can be seen from \eqref{L_bos_F} that the positivity of the kinetic terms (i.e. positivity of the expressions in the square brackets) is not guaranteed, and depends on the model. The model of our interest, as argued in Section \ref{sec_UV}, is determined by the choice
\begin{equation}
    {\cal G}=\tfrac{1}{1440}(X+\overbar X)^6~~\Longrightarrow~~{\cal G}^{(4)}(\phi_+)=\tfrac{1}{4}\phi^2_+=\phi_1^2~.\label{App_G}
\end{equation}
We then eliminate $F$ and $D$ by their equations of motion (these hold regardless of the choice of ${\cal G}$),
\begin{equation}
    F=-\tfrac{1}{2}m~,~~~D=0~,\label{App_F_eom}
\end{equation}
and the bosonic Lagrangian \eqref{L_bos_F} becomes
\begin{align}
\begin{split}
    {\cal L}_{\rm bos}=-(1+\tfrac{1}{2}\gamma m^2\phi_1^2)\partial\phi_1\partial\phi_1-\partial\phi_2\partial\phi_2&-\tfrac{1}{4}(1+\tfrac{1}{2}\gamma m^2\phi_1^2)F_{mn}F^{mn}\\
    &-\tfrac{1}{4}m^2-\tfrac{1}{16}\gamma m^4\phi_1^2~,
\end{split}
\end{align}
where the second line is the scalar potential of this model. As can be seen, for positive $\gamma$ the kinetic terms always have the correct sign for any value of $\phi_1$. At the same time, the scalar potential is stable, with the minimum at $\phi_1=0$, while $\phi_1$ has the mass $\sqrt{\gamma}m^2/4$.

As for the fermionic Lagrangian, up to one derivative and up to two fermions, it reads
\begin{align}
\begin{split}
    {\cal L}_{\rm fermi}= &-\tfrac{i}{2}\big[1-\gamma{\cal G}^{(4)}(m^2+mF+3m\overbar F+3F\overbar F+\tfrac{3}{2}D^2)\big]\chi\sigma^m\partial_m\bar\chi\\
    &-\tfrac{i}{2}\big[1-\gamma{\cal G}^{(4)}(2m\overbar F+3F\overbar F+\tfrac{3}{2}D^2)\big]\lambda\sigma^m\partial_m\bar\lambda\\
    &-\tfrac{\sqrt{2}}{4}\gamma m{\cal G}^{(4)}D(\chi\sigma^m\partial_m\bar\lambda-\lambda\sigma^m\partial_m\bar\chi)\\
    &+\tfrac{1}{2}\gamma{\cal G}^{(5)}\big[|m+F|^2\overbar F\chi^2+\tfrac{1}{2}(m+\overbar F)D^2\chi^2+(m+F)F\overbar F\lambda^2+\tfrac{1}{2}FD^2\lambda^2\\
    &\hspace{4.2cm}-\sqrt{2}i(m+F)\overbar FD\chi\lambda-\tfrac{i}{\sqrt{2}}D^3\chi\lambda\big]+{\rm h.c.}+\ldots~,\label{L_fermi_F}
\end{split}
\end{align}
After using Eqs. \eqref{App_G} and \eqref{App_F_eom} we have
\begin{equation}
    {\cal L}_{\rm fermi}=-\tfrac{i}{2}(1+\tfrac{1}{4}\gamma m^2\phi_1^2)(\chi\sigma^m\partial_m\bar\chi+\lambda\sigma^m\partial_m\bar\lambda)-\tfrac{1}{16}\gamma m^3\phi_1(\chi^2-\lambda^2)+{\rm h.c.}~.
\end{equation}
Similarly to the bosonic sector, fermions have the correct sign of the kinetic terms for any value of $\phi_1$ provided that $\gamma>0$. On the other hand, the masses of $\chi$ and $\lambda$ vanish at the minimum when $\phi_1=0$.

We conclude that the model has well-behaved kinetic terms, but possible contribution of higher-derivatives (such as $\Box\phi_1\Box\phi_1$) to the effective scalar potential may require further investigation.

\providecommand{\href}[2]{#2}\begingroup\raggedright\endgroup


\begin{thebibliography}{10}

\bibitem{Volkov:1973ix}
D.~V. Volkov and V.~P. Akulov, ``{Is the Neutrino a Goldstone Particle?},''
  \href{http://dx.doi.org/10.1016/0370-2693(73)90490-5}{{\em Phys. Lett. B}
  {\bfseries 46} (1973) 109--110}.

\bibitem{Rocek:1978nb}
M.~Rocek, ``{Linearizing the Volkov-Akulov Model},''
  \href{http://dx.doi.org/10.1103/PhysRevLett.41.451}{{\em Phys. Rev. Lett.}
  {\bfseries 41} (1978) 451--453}.

\bibitem{Ivanov:1978mx}
E.~A. Ivanov and A.~A. Kapustnikov, ``{General Relationship Between Linear and
  Nonlinear Realizations of Supersymmetry},''
  \href{http://dx.doi.org/10.1088/0305-4470/11/12/005}{{\em J. Phys. A}
  {\bfseries 11} (1978) 2375--2384}.

\bibitem{Lindstrom:1979kq}
U.~Lindstrom and M.~Rocek, ``{Constrained local superfields},''
  \href{http://dx.doi.org/10.1103/PhysRevD.19.2300}{{\em Phys. Rev. D}
  {\bfseries 19} (1979) 2300--2303}.

\bibitem{Ivanov:1982bpa}
E.~A. Ivanov and A.~A. Kapustnikov, ``{The nonlinear realization structure of
  models with spontaneously broken supersymmetry},''
  \href{http://dx.doi.org/10.1088/0305-4616/8/2/004}{{\em J. Phys. G}
  {\bfseries 8} (1982) 167--191}.

\bibitem{Samuel:1982uh}
S.~Samuel and J.~Wess, ``{A Superfield Formulation of the Nonlinear Realization
  of Supersymmetry and Its Coupling to Supergravity},''
  \href{http://dx.doi.org/10.1016/0550-3213(83)90622-3}{{\em Nucl. Phys. B}
  {\bfseries 221} (1983) 153--177}.

\bibitem{Casalbuoni:1988xh}
R.~Casalbuoni, S.~De~Curtis, D.~Dominici, F.~Feruglio, and R.~Gatto,
  ``{Nonlinear Realization of Supersymmetry Algebra From Supersymmetric
  Constraint},'' \href{http://dx.doi.org/10.1016/0370-2693(89)90788-0}{{\em
  Phys. Lett. B} {\bfseries 220} (1989) 569--575}.
  
\bibitem{Komargodski:2009rz}
Z.~Komargodski and N.~Seiberg, ``{From Linear SUSY to Constrained
  Superfields},'' \href{http://dx.doi.org/10.1088/1126-6708/2009/09/066}{{\em
  JHEP} {\bfseries 09} (2009) 066},
  \href{http://arxiv.org/abs/0907.2441}{{\ttfamily arXiv:0907.2441 [hep-th]}}.

\bibitem{Kuzenko:2010ef}
S.~M. Kuzenko and S.~J. Tyler, ``{Relating the Komargodski-Seiberg and
  Akulov-Volkov actions: Exact nonlinear field redefinition},''
  \href{http://dx.doi.org/10.1016/j.physletb.2011.03.020}{{\em Phys. Lett. B}
  {\bfseries 698} (2011) 319--322},
  \href{http://arxiv.org/abs/1009.3298}{{\ttfamily arXiv:1009.3298 [hep-th]}}.

\bibitem{Hughes:1986dn}
J.~Hughes and J.~Polchinski, ``{Partially Broken Global Supersymmetry and the
  Superstring},'' \href{http://dx.doi.org/10.1016/0550-3213(86)90111-2}{{\em
  Nucl. Phys. B} {\bfseries 278} (1986) 147--169}.

\bibitem{Hughes:1986fa}
J.~Hughes, J.~Liu, and J.~Polchinski, ``{Supermembranes},''
  \href{http://dx.doi.org/10.1016/0370-2693(86)91204-9}{{\em Phys. Lett. B}
  {\bfseries 180} (1986) 370--374}.

\bibitem{Ferrara:1995gu}
S.~Ferrara, L.~Girardello, and M.~Porrati, ``{Minimal Higgs branch for the
  breaking of half of the supersymmetries in N=2 supergravity},''
  \href{http://dx.doi.org/10.1016/0370-2693(95)01378-4}{{\em Phys. Lett. B}
  {\bfseries 366} (1996) 155--159},
  \href{http://arxiv.org/abs/hep-th/9510074}{{\ttfamily arXiv:hep-th/9510074}}.

\bibitem{Antoniadis:1995vb}
I.~Antoniadis, H.~Partouche, and T.~R. Taylor, ``{Spontaneous breaking of N=2
  global supersymmetry},''
  \href{http://dx.doi.org/10.1016/0370-2693(96)00028-7}{{\em Phys. Lett. B}
  {\bfseries 372} (1996) 83--87},
  \href{http://arxiv.org/abs/hep-th/9512006}{{\ttfamily arXiv:hep-th/9512006}}.

\bibitem{Ferrara:1995xi}
S.~Ferrara, L.~Girardello, and M.~Porrati, ``{Spontaneous breaking of N=2 to
  N=1 in rigid and local supersymmetric theories},''
  \href{http://dx.doi.org/10.1016/0370-2693(96)00229-8}{{\em Phys. Lett. B}
  {\bfseries 376} (1996) 275--281},
  \href{http://arxiv.org/abs/hep-th/9512180}{{\ttfamily arXiv:hep-th/9512180}}.

\bibitem{Bagger:1996wp}
J.~Bagger and A.~Galperin, ``{A New Goldstone multiplet for partially broken
  supersymmetry},'' \href{http://dx.doi.org/10.1103/PhysRevD.55.1091}{{\em
  Phys. Rev. D} {\bfseries 55} (1997) 1091--1098},
  \href{http://arxiv.org/abs/hep-th/9608177}{{\ttfamily arXiv:hep-th/9608177}}.

\bibitem{Ivanov:1997mt}
E.~A. Ivanov and B.~M. Zupnik, ``{Modified N=2 supersymmetry and
  Fayet-Iliopoulos terms},'' {\em Phys. Atom. Nucl.} {\bfseries 62} (1999)
  1043--1055, \href{http://arxiv.org/abs/hep-th/9710236}{{\ttfamily
  arXiv:hep-th/9710236}}.

\bibitem{Antoniadis:2017jsk}
I.~Antoniadis, J.-P. Derendinger, and C.~Markou, ``{Nonlinear $ \mathcal{N}=2 $
  global supersymmetry},''
  \href{http://dx.doi.org/10.1007/JHEP06(2017)052}{{\em JHEP} {\bfseries 06}
  (2017) 052}, \href{http://arxiv.org/abs/1703.08806}{{\ttfamily
  arXiv:1703.08806 [hep-th]}}.

\bibitem{Born:1934gh}
M.~Born and L.~Infeld, ``{Foundations of the new field theory},''
  \href{http://dx.doi.org/10.1098/rspa.1934.0059}{{\em Proc. Roy. Soc. Lond. A}
  {\bfseries 144} no.~852, (1934) 425--451}.

\bibitem{Cecotti:1986gb}
S.~Cecotti and S.~Ferrara, ``{Supersymmetric Born-Infeld Lagrangians},''
  \href{http://dx.doi.org/10.1016/0370-2693(87)91105-1}{{\em Phys. Lett. B}
  {\bfseries 187} (1987) 335--339}.

\bibitem{Rocek:1997hi}
M.~Rocek and A.~A. Tseytlin, ``{Partial breaking of global D = 4 supersymmetry,
  constrained superfields, and three-brane actions},''
  \href{http://dx.doi.org/10.1103/PhysRevD.59.106001}{{\em Phys. Rev. D}
  {\bfseries 59} (1999) 106001},
  \href{http://arxiv.org/abs/hep-th/9811232}{{\ttfamily arXiv:hep-th/9811232}}.

\bibitem{Bagger:1997pi}
J.~Bagger and A.~Galperin, ``{The Tensor Goldstone multiplet for partially
  broken supersymmetry},''
  \href{http://dx.doi.org/10.1016/S0370-2693(97)01030-7}{{\em Phys. Lett. B}
  {\bfseries 412} (1997) 296--300},
  \href{http://arxiv.org/abs/hep-th/9707061}{{\ttfamily arXiv:hep-th/9707061}}.

\bibitem{Gonzalez-Rey:1998vtf}
F.~Gonzalez-Rey, I.~Y. Park, and M.~Rocek, ``{On dual 3-brane actions with
  partially broken N=2 supersymmetry},''
  \href{http://dx.doi.org/10.1016/S0550-3213(99)00024-3}{{\em Nucl. Phys. B}
  {\bfseries 544} (1999) 243--264},
  \href{http://arxiv.org/abs/hep-th/9811130}{{\ttfamily arXiv:hep-th/9811130}}.

\bibitem{Ambrosetti:2009za}
N.~Ambrosetti, I.~Antoniadis, J.~P. Derendinger, and P.~Tziveloglou,
  ``{Nonlinear Supersymmetry, Brane-bulk Interactions and Super-Higgs without
  Gravity},'' \href{http://dx.doi.org/10.1016/j.nuclphysb.2010.03.027}{{\em
  Nucl. Phys. B} {\bfseries 835} (2010) 75--109},
  \href{http://arxiv.org/abs/0911.5212}{{\ttfamily arXiv:0911.5212 [hep-th]}}.

\bibitem{Wess:1992cp}
J.~Wess and J.~Bagger, {\em {Supersymmetry and supergravity}}.
\newblock Princeton University Press, Princeton, NJ, USA,
1992.
\newblock

\bibitem{Dudas:2017sbi}
E.~Dudas, S.~Ferrara, and A.~Sagnotti, ``{A superfield constraint for
  $\mathcal{N}=2\rightarrow\mathcal{N}=0$ breaking},''
  \href{http://dx.doi.org/10.1007/JHEP08(2017)109}{{\em JHEP} {\bfseries 08}
  (2017) 109}, \href{http://arxiv.org/abs/1707.03414}{{\ttfamily
  arXiv:1707.03414 [hep-th]}}.

\bibitem{Kuzenko:2017gsc}
S.~M. Kuzenko and G.~Tartaglino-Mazzucchelli, ``{New nilpotent ${\cal N}= 2$
  superfields},'' \href{http://dx.doi.org/10.1103/PhysRevD.97.026003}{{\em
  Phys. Rev. D} {\bfseries 97} no.~2, (2018) 026003},
  \href{http://arxiv.org/abs/1707.07390}{{\ttfamily arXiv:1707.07390
  [hep-th]}}.

\bibitem{Cribiori:2016hdz}
N.~Cribiori, G.~Dall'Agata, and F.~Farakos, ``{Interactions of N Goldstini in
  Superspace},'' \href{http://dx.doi.org/10.1103/PhysRevD.94.065019}{{\em Phys.
  Rev. D} {\bfseries 94} no.~6, (2016) 065019},
  \href{http://arxiv.org/abs/1607.01277}{{\ttfamily arXiv:1607.01277
  [hep-th]}}.

\bibitem{Aldabergenov:2021obf}
Y.~Aldabergenov, A.~Chatrabhuti, and H.~Isono, ``{Nilpotent superfields for
  broken abelian symmetries},''
  \href{http://dx.doi.org/10.1140/epjc/s10052-021-09320-4}{{\em Eur. Phys. J.
  C} {\bfseries 81} no.~6, (2021) 523},
  \href{http://arxiv.org/abs/2103.11217}{{\ttfamily arXiv:2103.11217
  [hep-th]}}.

\bibitem{Fayet:1975yi}
P.~Fayet, ``{Fermi-Bose Hypersymmetry},''
  \href{http://dx.doi.org/10.1016/0550-3213(76)90458-2}{{\em Nucl. Phys. B}
  {\bfseries 113} (1976) 135}.

\end{thebibliography}
\end{document}